\documentclass[fleqn,usenatbib]{mnras}

\usepackage{newtxtext,newtxmath}

\usepackage[T1]{fontenc}

\DeclareRobustCommand{\VAN}[3]{#2}
\let\VANthebibliography\thebibliography
\def\thebibliography{\DeclareRobustCommand{\VAN}[3]{##3}\VANthebibliography}

\usepackage{graphicx}	
\usepackage{amsmath}	
\usepackage{amssymb}	
\usepackage{booktabs}   
\usepackage{multirow}   
\usepackage{comment}    
\usepackage{gensymb}    
\usepackage{subfig}    

\title[Metallicity dependence on local and global quantities]{The metallicity's fundamental dependence on both local and global galactic quantities}

\author[W. M. Baker et al.]{
William M. Baker$^{1,2}$\thanks{E-mail: wb308@cam.ac.uk},
Roberto Maiolino$^{1,2,3}$,
Francesco Belfiore$^{4}$,
Mirko Curti$^{1,2}$,
Asa F. L. Bluck$^{5}$,
\newauthor{}
Lihwai Lin$^{6}$,
Sara L. Ellison$^{7}$,
Mallory Thorp$^{7}$,
Hsi-An Pan$^{8}$
\\
$^{1}$Kavli Institute for Cosmology, University of Cambridge, Madingley Road, Cambridge, CB3 0HA, UK\\
$^{2}$Cavendish Laboratory - Astrophysics Group, University of Cambridge, 19 JJ Thomson Avenue, Cambridge, CB3 0HE, UK\\
$^{3}$Department of Physics and Astronomy, University College London, Gower Street, London WC1E 6BT, UK\\
$^{4}$INAF— Osservatorio Astrofisico di Arcetri, Largo E. Fermi 5, I-50125, Florence, Italy\\
$^{5}$Department of Physics, Florida International University, 11200 SW 8th Street, Miami, FL, USA\\
$^{6}$Institute of Astronomy \& Astrophysics, Academia Sinica, Taipei 10617, Taiwan\\
$^{7}$Department of Physics \& Astronomy, University of Victoria, Finnerty Road, Victoria, British Columbia, V8P 1A1, Canada \\
$^{8}$Department of Physics, Tamkang University, No.151, Yingzhuan Road, Tamsui District, New Taipei City 251301, Taiwan 
}

\date{Accepted XXX. Received YYY; in original form ZZZ}

\pubyear{2015}

\begin{document}
\label{firstpage}
\pagerange{\pageref{firstpage}--\pageref{lastpage}}
\maketitle

\begin{abstract}
We study the scaling relations between gas-phase metallicity, stellar mass surface density ($\Sigma _*$), star formation rate surface density ($\Sigma _{SFR}$), and molecular gas surface density ($\Sigma_{H_2}$) in local star-forming galaxies on scales of a kpc.
We employ optical integral field spectroscopy from the MaNGA survey, and ALMA data
for a subset of MaNGA galaxies. We use Partial Correlation Coefficients and Random Forest regression to determine the relative importance of
local and global galactic properties in setting the gas-phase metallicity. We find that the local metallicity depends primarily on $\Sigma _*$ (the resolved mass-metallicity relation, rMZR), and has a secondary anti-correlation with $\Sigma _{SFR}$ (i.e. a spatially-resolved version of the `Fundamental Metallicity Relation', rFMR). We find that $\Sigma_{H_2}$  is less important than $\Sigma_{SFR}$ in determining the local metallicity. This result indicates that gas accretion, resulting in local metallicity dilution and local boosting of star formation, is unlikely to be the primary origin of the rFMR. 
The local metallicity depends {\it also} on the global properties of galaxies. We find a strong dependence on the total stellar mass ($M_*$) and a weaker (inverse) dependence on the total SFR. The global metallicity scaling relations, therefore, do not simply stem out of their resolved counterparts; global properties and processes, such as the global gravitational potential well, galaxy-scale winds and global redistribution/mixing of metals, likely contribute to the local metallicity, in addition to local production and retention.

\end{abstract}




\section{Introduction}

The gas-phase metallicity of local galaxies has been observed to correlate strongly with stellar mass following the so-called mass-metallicity relation (MZR,  \citealt{2004Tremonti}, for a review \citealt{Maiolino2019}). As the stellar mass increases, the metallicity increases, until the relation flattens at high stellar masses. The MZR has been observed to evolve significantly with redshift, in the sense that high-redshift galaxies tend to be more metal poor than local ones at a given stellar mass \citep[e.g.][]{Maiolino2008A&A...488..463M, Troncoso2014A&A...563A..58T, Sanders2021ApJ...914...19S}.

The MZR has been show to have additional dependencies on other galactic properties such as specific star-formation rate (sSFR=SFR/$M_\star$) and half-light radius, as found in \citet[][]{2008Ellison}.
\citet{2010Mannucci} confirmed via more detailed analysis that the MZR has a secondary dependence on the star-formation rate (SFR). At a given stellar mass, galaxies that have higher star formation rates have lower metallicity. A common scenario to interpret this metallicity-SFR anticorrelation is that accretion of nearly pristine (or metal-poor) gas both dilutes the metallicity and provides fuel for more star formation \citep[e.g.][]{Lilly2013,Peng2014FromHaloes,Dave2011MNRAS.416.1354D,Torrey2018MNRAS.477L..16T}. \citet{2010Mannucci}  also found that, up to $z\sim2.5$, the three-dimensional $M_\star$-Z-SFR relationship does not appear to evolve with redshift. They named this relation the `Fundamental Metallicity Relation' (FMR). They showed that the appearance of the evolution of the MZR with redshift was a consequence of higher redshift galaxies having greater SFR, thereby populating a different region of the FMR.

The existence of an FMR has been questioned in a few papers, 
such as \citet{2013SanchezA&A...554A..58S,Sanchez2017MNRAS.469.2121S, Barrera_NO_FMR_2017ApJ...844...80B}.
However, further support for the existence of the FMR was given by the work of \cite{Salim2014ApJ...797..126S}. They reanalysed previous datasets which had found no evidence of the FMR and found that, independently of the metallicity indicators they used, metallicity was anti-correlated with sSFR for galaxies with stellar masses up to $10^{10.5} \rm M_\odot$. \cite{Cresci2019A&A...627A..42C} further reanalysed data from the Calar Alto Legacy Integral Field Area (CALIFA) survey and the Mapping Nearby Galaxies at Apache Point Observatory (MaNGA) survey, focusing on resolving discrepancies with results that do not find evidence for an FMR \citep[such as][]{2013SanchezA&A...554A..58S,Sanchez2017MNRAS.469.2121S, Barrera_NO_FMR_2017ApJ...844...80B}. They confirm that, when observations are used in a consistent way and uncertainties are taken into account, the data show evidence for the existence of an FMR which does not evolve out to $z \sim 2.3$.

First evidence for a resolved FMR was found in \cite{Maiolino2019}, \cite{Xi2021arXiv}, \cite{Sanchez-Menguiano2019} and \citet{Schaefer_N/O_2022ApJ...930..160S}. In the latter work they investigate the Nitrogen to Oxygen relation and how it relates to metallicity, stellar mass and star formation rate. In addition to their primary analysis, they find an inverse relation between metallicity and resolved star formation rate, and a positive relation between metallicity and resolved stellar mass.

As mentioned, the suggested physical mechanisms behind the MZR (and hence likely the FMR) are often based upon so-called gas-regulator models \citep[such as][]{Lilly2013,Peng2014FromHaloes}. At their most simple, these mechanisms are one, or both of, pristine gas accretion (metal poor inflow), or some form of enriched (metal-loaded) outflow. This suggests that the secondary dependence of the MZR on SFR could be indirectly tracing gas content via the Schmidt-Kennicutt law \citep{Schmidt1959, Kennicutt1998}, that connects SFR and cold gas content.

A secondary dependence of the MZR with gas mass is indeed found for atomic gas in \cite{Bothwell2013MNRAS.433.1425B}, \cite{Brown2018MNRAS.473.1868B} and \cite{Chen_HI_2022arXiv220508331C}. 
\cite{Bothwell2016MNRAS.455.1156B} found that both molecular gas mass and total gas mass correlate negatively with the offset from the MZR. Using principal component analysis (PCA) they found that the secondary dependence of the MZR with gas mass is stronger than that with SFR, supporting the hypothesis that SFR is probably only a tracer for gas mass in the FMR.

Simulations also predict the presence of MZR with a secondary dependence on SFR (or indirectly via gas content); \citep{Torrey2019MNRAS.484.5587T, Trayford_and_Joop_2019MNRAS.485.5715T, DeLucia2020MNRAS.498.3215D, VanLoon2021MNRAS.504.4817V}.
\cite{Torrey2019MNRAS.484.5587T} used the IllustrisTNG simulation suite to make predictions for the MZR across a large redshift range. They found evidence for a secondary dependence on SFR  in their simulation, and they ascribe the effect to higher gas fractions, rather than variations in the ability to retain metals. 
\cite{VanLoon2021MNRAS.504.4817V} used the EAGLE hydrodynamical simulation, finding that the scatter of the MZR is affected by gas fraction, inflow rate and outflow rate. 
\cite{DeLucia2020MNRAS.498.3215D} used the GAEA semi-analytic model to analyse the MZR. They found that cold gas content is the quantity driving the apparent evolution of the MZR over redshift. They show that as the amount of gas decreases, which is a consequence of both gas consumption and a lack of fresh gas supply, the gas metallicity increases (explaining offsets above the MZR), whilst as the gas content increases, the metallicity decreases due to dilution (explaining offsets below the MZR).

The advent of extensive Integral Field Spectroscopy (IFS) surveys has allowed various teams to establish that the
MZR is also  observed on spatially resolved scales ($\sim$kpc), i.e. there exists a relationship between stellar mass surface density $\Sigma_*$ and local metallicity \citep[e.g.][]{RosalesOrtega2012ApJ...756L..31R,Barrera_global_from_local_radial2016MNRAS.463.2513B}. This has been interpreted as the local gravitational potential of the disc being effective in retaining metals, and it has been  suggested that the global MZR stems out of the its locally resolved version \citep[][]{Almeida_and_Menguiano2019ApJ...878L...6S}.

For what concerns the FMR, the existence of a spatially resolved version (i.e. a secondary dependence on $\Sigma_{SFR}$) is still a matter of discussion, with some studies finding evidence supporting it \citep[][]{Maiolino2019,Xi2021arXiv,Sanchez-Menguiano2019}, whilst others \citep{Barrera_NO_FMR_2017ApJ...844...80B} found no evidence for its existence. These works adopt different approaches in identifying the potential secondary dependence on $\Sigma_{SFR}$ (e.g. exploring the change of the metallicity in bins of $\Sigma _{SFR}$, as in \cite{Maiolino2019}, versus  investigating the metallicity dispersion when using the SFR as additional parameter to describe the MZR).

Within this context, one should take into account that the quantities involved in the FMR, and in particular stellar mass and SFR, are also linked through several scaling laws on both global and resolved scales. For instance, as already mentioned, SFR and molecular gas mass (i.e. the fuel for star formation) are correlated via the the so-called Schmidt-Kennicutt (SK) relation \citep{Schmidt1959, Kennicutt1998} which is also valid on resolved scales as the rSK \citep[][]{LeroySFE_2008AJ....136.2782L}. The Molecular Gas Main Sequence (MGMS) \citep{Lin2019} links stellar mass and molecular gas mass (rMGMS on resolved scales). Finally, the Star Forming Main Sequence \citep[SFMS,][]{2004MNRASBrinchmann,Renzini+PengMS2015ApJ...801L..29R}, whilst not a fundamental scaling relation \citep[][]{Lin2019, Ellison2021AlmaQuest5,Morselli2020MNRAS.496.4606M,Baker2022MNRAS.510.3622B}, describes the observed correlation between the star formation rate of a galaxy and its stellar mass, on both resolved (rSFMS) and global scales. Therefore, the correlation of metallicity with each of these quantities may actually not be a direct, fundamental relation, but a by product of indirect correlations.

Some of the open questions we want to answer in this paper are: does the resolved FMR exist? Is the global MZR simply a product of the local MZR? Are radial metallicity gradients simply a consequence of the local, resolved MZR and of the radially decreasing $\Sigma _*$ profile? Is the secondary dependence of metallicity on SFR, if any, a consequence of a dependence on molecular gas mass? Or does the metallicity primarily depend on a different set of quantities?

We start by reviewing the surveys and data products that we use, then we explore the effect of varying selection criteria (Sec. \ref{sec:data}).
We introduce the two main techniques use in our analysis: 1) the partial correlation coefficients, which allow us to measure correlations between quantities where some are intrinsically inter-correlated, and others indirectly correlated; 2) the Random Forest regression analysis, which allows us to explore (simultaneously) non-linear and non-monotonic intrinsic correlations within our data. 
We first explore our results for the MaNGA sample \citep[][]{Bundy2015}, where we investigate the different properties and dependencies of the spatially resolved MZR (rMZR) and  of the spatially resolved FMR (rFMR).
We then explore the subset of MaNGA galaxies that are part of the ALMA-MaNGA QUEnching and STar formation  \citep[ALMaQUEST][]{Lin2020} sample, for which we have spatially resolved CO data, enabling us to probe the effect of molecular gas on the rMZR and rFMR. 

 We assume $\rm H_0=70$km $\text{s}^{-1}$ $\text{Mpc}^{-1}$, $\Omega_m=0.3$ and $\Omega_{\Lambda}=0.7$ and a \cite{Kroupa_IMF_2001MNRAS.322..231K} Initial Mass Function throughout this paper.

\section{Data and selection criteria}
\label{sec:data}

\subsection{MaNGA}
The primary source of data we use is from Data Release (DR) 15 of the MaNGA (Mapping Nearby Galaxies at Apache Point Observatory) survey, which is part of the fourth generation Sloan Digital Sky Survey \citep[SDSS,][]{Blanton2017AJ....154...28B}. We use this data release as it was the only publicly available data at the time of our analysis.

The MaNGA survey \citep{Bundy2015} obtained integral field spectroscopy for $\sim$ 10,000 local ($z\sim0.03$) galaxies. It selects galaxies to follow a roughly flat distribution in $i$-band luminosity (as a proxy for stellar mass). MaNGA covers galaxies out to 1.5 (2/3 of the sample) or 2.5 (1/3 of the sample) effective radii with its spatial resolution elements (spaxels). MaNGA galaxies are typically sampled with  about 3 to 7 radial spatial resolution elements.
The point spread function (PSF) full-width-half-maximum (FWHM) is approximately 2.5''.
Details on the sample selection, data reduction, flux calibration and data quality are presented in \citet{Wake2017AJ....154...86W,Law2016AJ....152...83L,Yan2016AJ....151....8Y, Yan2016AJ....152..197Y} respectively.

We use a combination of the MaNGA DAP  \citep[][]{BelfioreDAP2019AJ....158..160B, Westfall2019AJ....158..231W} data alongside PIPE3D \citep[][]{Sanchez2016, SanchezPipe3d2016RMxAA..52..171S} data products. 
We take resolved data for: emission line fluxes, stellar masses, equivalent widths, and galactocentric distance.
Stellar masses per pixel, integrated stellar masses, mass-weighted age of the stellar population (at the effective radius) and integrated star formation rates  (via H$\alpha$) are obtained from the Pipe3D pipeline \citep[][]{Sanchez2016,SanchezPipe3d2016RMxAA..52..171S}.
We note that the Pipe3D data products use a Salpeter \citep[][]{Salpeter_IMF_1955ApJ...121..161S} initial mass function (IMF) for determining the stellar masses and SFRs, and, as we use the \citet{Kennicutt2012ARA&A..50..531K} calibration for our resolved SFR, we convert the Pipe3D stellar masses and SFRs to a Kroupa \citep[][]{Kroupa_IMF_2001MNRAS.322..231K} IMF. This is outlined in more detail in section \ref{sec: deriving quantities}.

The emission lines are obtained from the MaNGA DAP \citep[][]{BelfioreDAP2019AJ....158..160B, Westfall2019AJ....158..231W}. We use H$\alpha$, H$\beta$, [OIII]$\lambda$5007, [OII]$\lambda$3728, [NII]$\lambda$6584 and [SII]$\lambda$6717,31. The DAP uses a hybrid binning approach using pPXF \citep[penalized pixel-fitting][]{CappellaripPXF2017MNRAS.466..798C} where voronoi binning \citep[][]{Cappellari_voronoi_2003MNRAS.342..345C} is used for the continuum (where it aims for a homogeneous S/N ratio) and individual spaxels for the emission lines, which allows it to fit both the continuum and emission lines at the same time. 

The galaxy redshifts and inclinations are taken from the NASA Sloan Atlas\footnote{http://nsatlas.org/data}.

\subsection{ALMaQUEST}
The ALMaQUEST survey \citep{Lin2020} is designed to obtain information on the molecular gas distributions via CO(1-0) ALMA maps for 46 MaNGA galaxies, spanning a broad interval in terms of distance from the main sequence.  Targets were selected to have a broad range of specific star formation rates (sSFR), spanning from galaxies in the green valley (22 of them), on the Main Sequence (12 of them) and above the Main Sequence (12 starburst galaxies). The galaxies in the sample have stellar masses between $10^{10}~M_{\odot}$ and $10^{11.5}~M_{\odot}$.  ALMaQUEST data has an angular resolution of approximately 2.5'', which nicely matches the MaNGA PSF. The data spans a range in CO line flux sensitivity spanning from 0.02 to 0.1 Jy km s$^{-1}$ beam$^{-1}$.
For our data we get a minimum value of $\rm \Sigma_{H_2}\sim6.5\, M_\odot/kpc^2$ and a maximum of $\rm \Sigma_{H_2}\sim9.4\, M_\odot/kpc^2$.  More information about the ALMaQUEST survey can be found in \cite{Lin2020}.

\subsection{Selection of galaxies and spaxels}

\label{sec:selection of galaxies}
Interacting and merging galaxies undergo complex mixing processes and show an irregular distribution of stars, gas and metals. As in this work we focus on more regular and secular evolving galaxies, we remove interacting pairs and post-merger galaxies from the sample using a catalog based on visual inspection of SDSS images for the MaNGA sample (Mallory Thorp, private comm.).
We retain galaxies that show no disturbance but contain another galaxy visible in the IFU, in order to avoid an overly stringent cut. Features that indicate a merger event include tidal tales, rings and other highly perturbed morphological indicators.

In order to avoid extreme extinction correction, which may not be recovered properly by the Balmer decrement, and also to avoid strong inclination corrections, we remove highly inclined galaxies by applying an inclination cut of $\frac{b}{a}\geq0.35$.

We do not pre-select passive or star forming galaxies a priori, but select galactic regions on a spaxel to spaxel basis, across all galaxies matching the preliminary requirements discussed above. More specifically,
we select star-forming regions using a [NII]/H$\alpha$ versus [OIII]/H$\beta$ Baldwin-Philips-Terlevich (BPT); \citep{Baldwin} diagnostic diagram, utilising the \citet{2003Kauffmann} dividing line. Selecting star-forming regions with such criteria is required as the metallicity tracers that we use (see further below) are only calibrated for these regions.

Although we do not pre-select galaxies based on their spectral classification or based on the position on the M$_*$--SFR or color-magnitude diagrams, as a matter of fact passive galaxies
are essentially excluded by the spaxels selection requirements as well as the requirements on the H$\alpha$ signal-to-noise and EW outlined below.
However, many star forming regions in green valley galaxies are retained by our selection \citep[see][where they show that there are a considerable number of star forming spaxels in the ALMaQUEST green valley sample]{LinGV2022ApJ...926..175L}.

Following the analysis in  \cite{2010Mannucci} our baseline for the MaNGA sample is to use a S/N cut of 25 on H$\alpha$ (which represents essentially a cut in $\Sigma _{SFR}$, meaning that we do not explore very quiescent regions).  This automatically ensures that all relevant metal lines required for the metallicity diagnostics are detected with S/N$>$3 without introducing a bias in the metallicity determination. 
We use the same criteria when exploring the ALMaQUEST galaxies although with the addition of a S/N cut of 3 on the CO(1-0) emission line.

In addition to S/N cuts, we apply a cut on the equivalent width for H$\alpha$ of EW(H$\alpha\geq6$\AA), in order to minimise contamination from diffuse ionised gas (DIG). We also test an equivalent width cut of 14\AA\,, for which H$\alpha$ is identified as being due to star formation with little to no contribution from DIG on kpc scales \citep[][]{Lacerda2018MNRAS.474.3727L} and find that it makes no significant difference to any of the conclusions of this paper \citep[as expected by the results of][]{Mannucci_DIG_2021MNRAS.508.1582M}.

These selection criteria result into a total of about 1.12$\times 10^6$ spaxels from 2002 galaxies. However, as the spaxels (0.5$''$ in size) in the MaNGA maps oversample the survey PSF (2.5$''$) by a large factor (about a factor of $\sim$20 in area),
the number of {\it independent} regions is actually about 56,000.

The statistics for the ALMaQUEST sample is obviously much smaller: $\sim$25770 spaxels matching the galaxy selection and optical nebular line criteria discussed above, which (as a consequence of oversampling) correspond to about $\sim$ 1290 independent regions. These numbers are further reduced for the subset of spaxels that have CO detection:
giving a total of  13,450 spaxels, corresponding to about 670 independent spaxels.  We note that the CO detected spaxels are, on average, closer to the centre of the galaxies than the undetected spaxels (with median distance of 0.95$r_e$, and 1.53$r_e$ respectively).

\subsection{Deriving physical quantities}
\label{sec: deriving quantities}
All emission lines are corrected for extinction by using the H$\alpha$/H$\beta$ Balmer decrement, assuming an intrinsic value of 2.86, and adopting a 
 \citet{Cardelli1989} Milky Way-like extinction curve with $R_v$=3.1. Results do not change significantly by assuming a \citet{Calzetti2000} attenuation curve.

The star-formation rate surface density is calculated using the dust-corrected H$\alpha$ flux via the relation \citep[][for a Kroupa IMF]{Kennicutt2012ARA&A..50..531K}
\begin{equation}
    \mathrm{log(SFR\, /M_\odot yr^{-1})}=\mathrm{log(L_{\rm H\alpha}\, /erg\,s^{-1})}-41.27.
\end{equation}
We convert the integrated SFR, obtained via PIPE3D for a Salpeter IMF with the \citet{1998KennicuttSFR} calibration, and the integrated stellar mass to the new calibration via the conversion given in \citet{Kennicutt2012ARA&A..50..531K}:
\begin{equation}
    \rm SFR_{K2012}= 0.68\,SFR_{K1998}
\end{equation}

The resolved stellar masses and star formation rates are converted into surface densities by dividing each spaxel quantity by the spaxel area.

To calculate the metallicity, we simultaneously use a combination of nine strong-line diagnostics tied to the T$_e$ temperature scale. These diagnostics are R23, R2, R3, O32, N2, S2, O3S2, O3N2, and R3S2, which are defined in Table \ref{table:cali}, in combination with the empirical calibrations given in \citet{2020CurtiMNRAS.491..944C}. This enables us to use multiple diagnostics at the same time which means that we can break any degeneracies associated with diagnostics that are double valued, whilst also incorporating diagnostics that are most effective at different ranges of metallicities \citep[see ][ for details]{2020CurtiMNRAS.491..944C}.  Each spaxel's metallicity is found by minimizing the chi-square defined by the observed values and the values predicted by the calibrations. It also takes into account uncertainties on the observed line ratio and the intrinsic dispersion of the calibration. It does this via Markov Chain Monte Carlo (MCMC), enabling it to vary line fluxes based upon an assumed Gaussian distribution \citep[For more details on the procedure see][]{2017MNRASCurti, 2020CurtiMNRAS.491..944C}.

\begin{table}
\caption{Metallicity diagnostics used in this paper - same calibrations as in \citet{2020CurtiMNRAS.491..944C}.}
\centering
\label{table:cali}
\begin{tabular}{l c c c c}
\toprule
\multirow{2}{*}{Tracer} & \multirow{2}{*}{Definition} \\[+8pt]

\midrule

R23 & $ \text{log}\big(\frac{\text{[OIII]}\lambda5007,4958+\text{[OII]}\lambda3727}{H\beta}\big) $ \\[+10pt]

$\text{O32}$ & $\text{log}\big(\frac{\text{[OIII]}\lambda5007}{\text{[OII]}\lambda3727}\big)$  \\[+10pt]

$\text{R2}$ & $ \text{log}\big(\frac{\text{[OII]}\lambda3727}{H\beta}\big)$  \\[+10pt]

$\text{R3}$ & $\text{log}\big(\frac{\text{[OIII]}\lambda5007}{H\beta}\big)$  \\[+10pt]

N2 & $\text{log}\big(\frac{\text{[NII]}\lambda6584}{H\alpha}\big) $ \\[+10pt]

S2 & $ \text{log}\big(\frac{\text{[SII]}\lambda6718,32}{H\alpha}\big)$  \\[+10pt]

O3N2 & $\text{log}\big(\frac{\text{[OIII]}\lambda5007/H\beta}{\text{[NII]}\lambda6584/H\alpha}\big)$   \\[+10pt]

O3S2 & $\text{log}\big(\frac{\text{[OIII]}\lambda5007/H\beta}{\text{[SII]}\lambda6718,32/H\alpha}\big)$  \\[+10pt]

R3S2 & $\text{log}\big(\frac{\text{[OIII]}\lambda5007}{H\beta}+\frac{\text{[SII]}\lambda6718,32}{H\alpha}\big)$  \\[+10pt]
\bottomrule
\end{tabular}
\end{table}

For the ALMaQUEST sample of galaxies we convert the CO luminosity (obtained from the flux via the definition listed in \citealt{Solomon1997}) into the molecular gas mass density in two different ways. The simplest approach is to use a Milky Way like constant conversion factor of $\alpha_{\rm CO}=4.35\, \text{M}_\odot (\text{K km/s pc}^2)^{-1}$ \citep{Bolatto2013}. However, the conversion factor is known to depend steeply on the gas metallicity. Different parameterisations of this metallicity dependence have been proposed in the past \citep[see ][ for a review]{Bolatto2013}. In this paper we adopt the relation obtained by the extensive study presented in \cite{Accurso2017MNRAS.470.4750A}:
\begin{equation}
  \alpha_{\rm CO}=4.35\bigg(\frac{Z}{Z_\odot}\bigg)^{-1.6} \text{M}_\odot (\rm K\, km\,s^{-1} pc^2)^{-1}.  
\end{equation}
We will discuss the impact of the conversion factor on the overall results.

Finally, we correct all surface densities for inclination via the equation
\begin{equation}
    \Sigma_{\text{\rm corrected}}=\bigg(\frac{b}{a}\bigg)\,\Sigma_{\text{\rm observed}}.
\end{equation}

\begin{figure}
    \centering
    \includegraphics[width=\columnwidth]{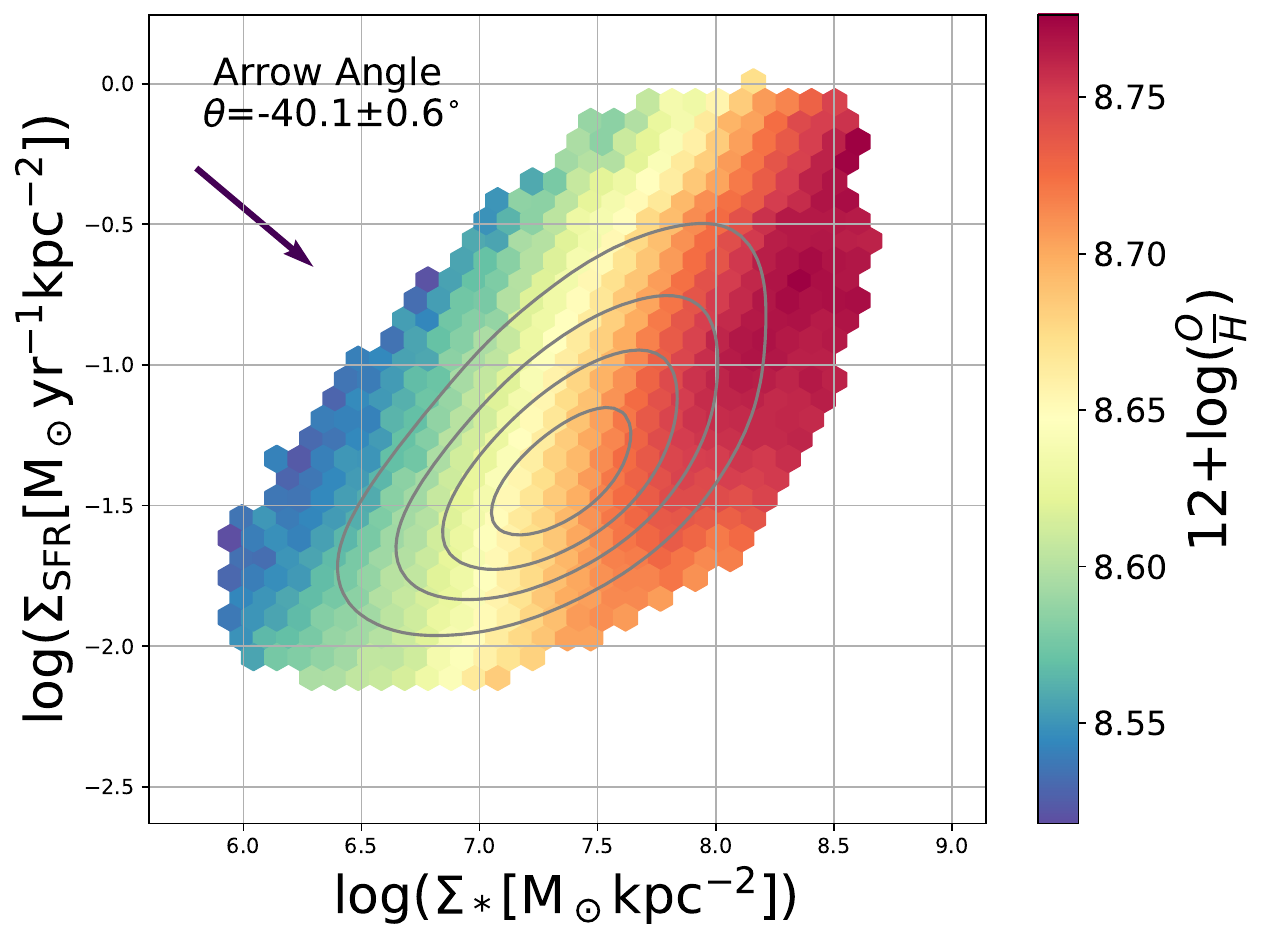}
    \caption{Star formation rate surface density versus stellar mass surface density colour coded by metallicity. The colour of each hexagon gives the mean metallicity of the spaxels contained in it. The arrow angle gives direction of the steepest gradient across the entire distribution, as obtained by the Partial Correlation Coefficients (see Sec. \ref{sec:methodology}) and is measured from the positive x-axis direction. Both the colour shading and the arrow unambiguously show the presence of a resolved FMR, i.e. a primary dependence of the metallicity on $\Sigma _*$ and a secondary (inverse) dependence on $\Sigma _{SFR}$.
    The contours indicate the density distribution of spaxels in the diagram, where the outer density contour encloses 90\% of the spaxels. }
    \label{fig:resolved_FMR_manga}
\end{figure}

\begin{figure}
    \centering
    \includegraphics[width=\columnwidth]{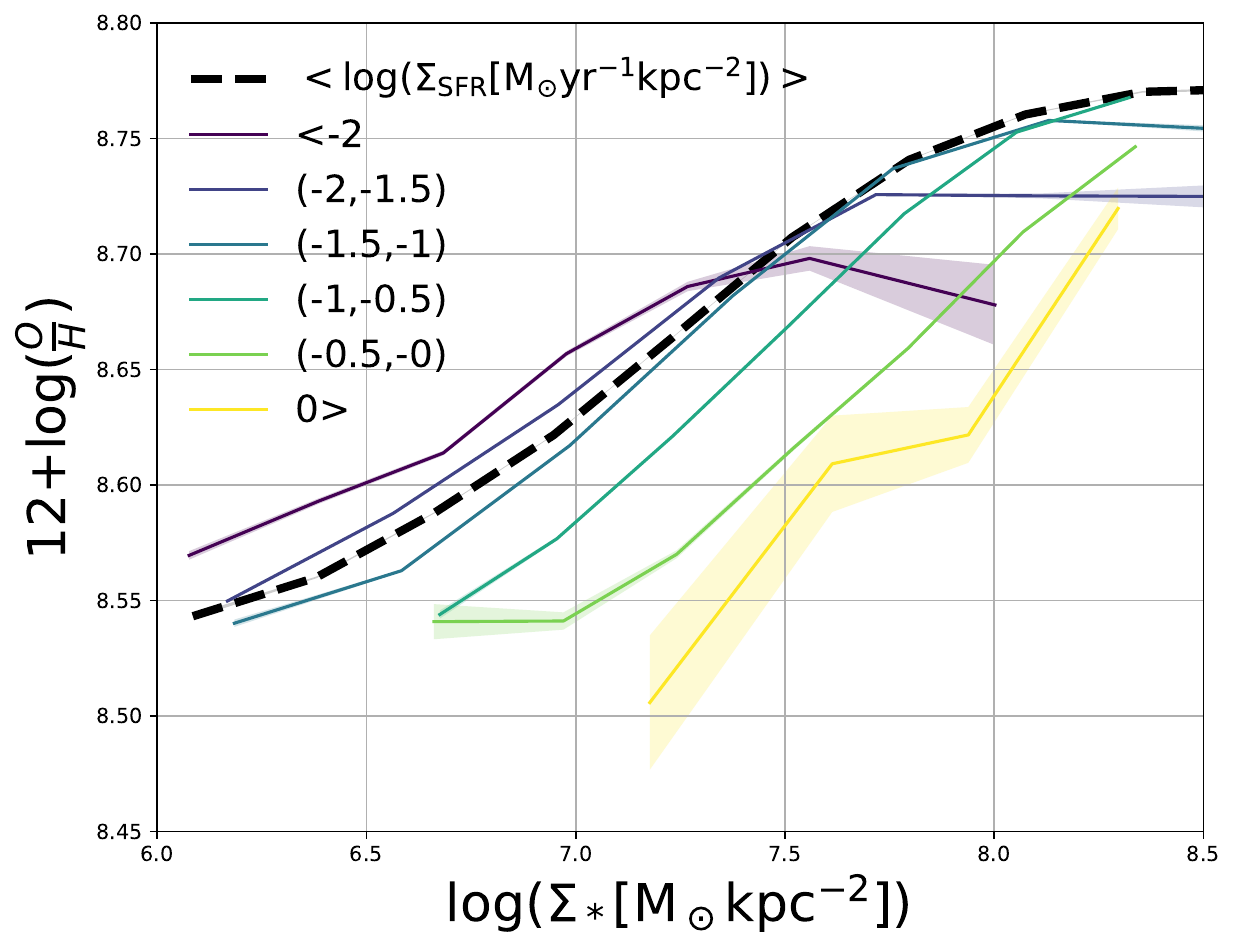}
    \caption{Metallicity versus stellar mass surface density binned by tracks of the star formation rate surface density.
    The shading corresponds to the standard error on the mean of the metallicity for each bin. The black dashed line represents the track for the data in the case of no binning in $\Sigma_{\rm SFR}$.
    Clearly (except for regions at high $\Sigma _*$ and low $\Sigma _{SFR}$) at a given $\Sigma _*$ the metallicity decreases at higher $\Sigma _{SFR}$ i.e. confirming the presence of the rFMR.}
    \label{fig:FMR_tracks}
\end{figure}

\begin{figure*}
    \centering
    \includegraphics[width=0.8\paperwidth]{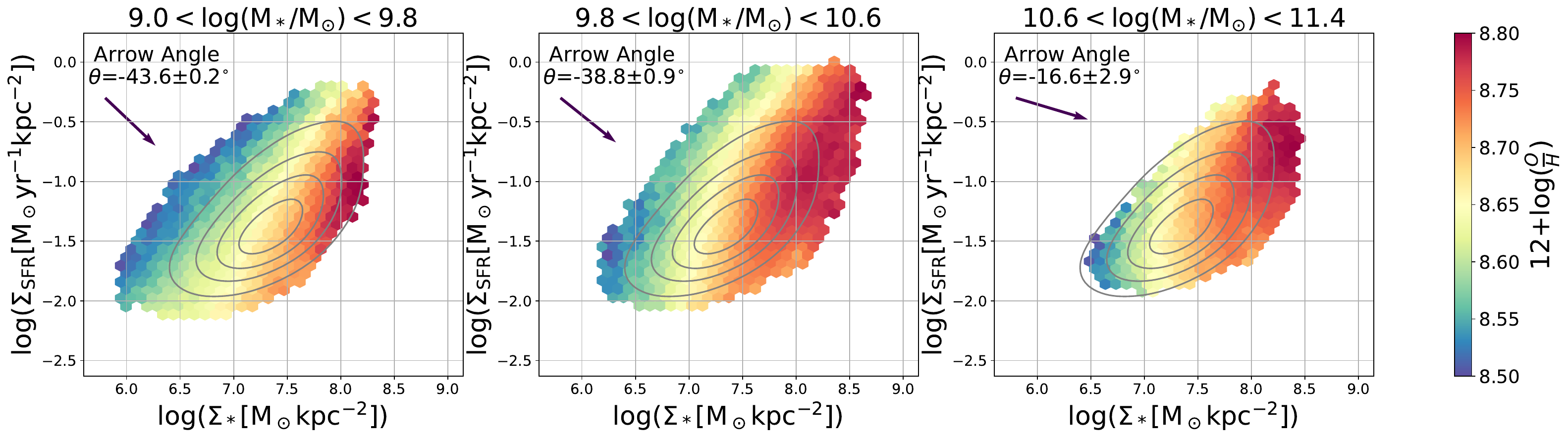}
    \caption{Same as Figure \ref{fig:resolved_FMR_manga}, but where the galaxies are divided in three bins of total stellar mass, $M_*$, as indicated on the top of each panel. The figure illustrates that the local metallicity has also a dependence on total stellar mass and that the inverse correlation with $\Sigma _{SFR}$ is strongest for low mass galaxies.
    As a common reference, the density contours are the same as for the total population in Figure \ref{fig:resolved_FMR_manga}. }
    \label{fig:multiplot}
\end{figure*}

\begin{figure*}
    \centering
    \includegraphics[width=0.8\paperwidth]{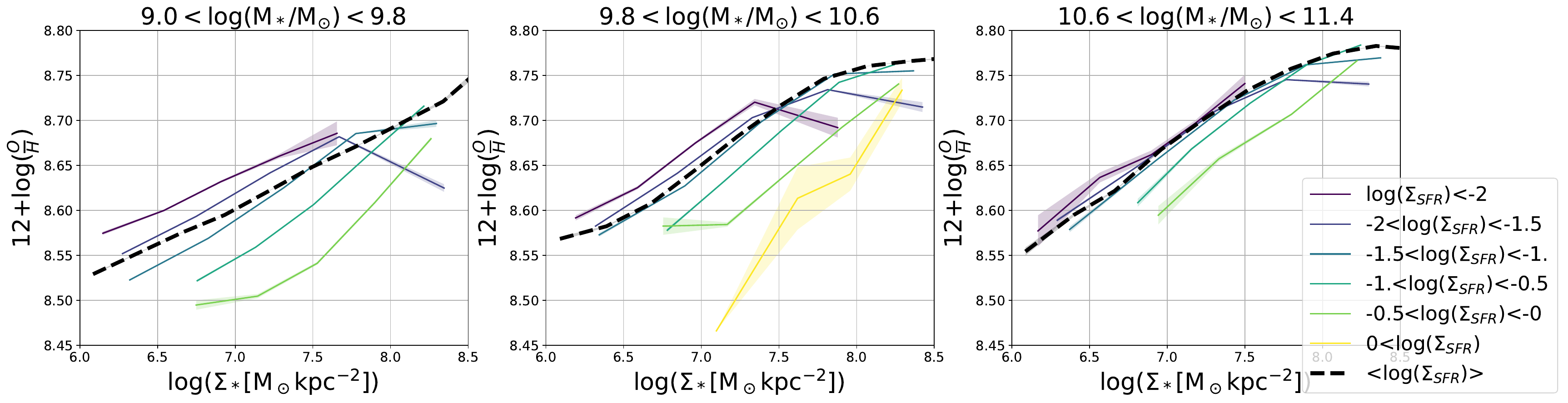}
    \caption{Same as Figure \ref{fig:FMR_tracks}, but where the galaxies are divided in three bins of total stellar mass, $M_*$, as indicated on the top of each panel. The figure further illustrates that the local metallicity has also a dependence on total stellar mass and that the inverse correlation with $\Sigma _{SFR}$ is strongest for low mass galaxies
    (as it is the case for the inverse dependence on SFR for the global FMR). }
    \label{fig:multiplot_tracks}
\end{figure*}

\section{Methodology}

\label{sec:methodology}
We use two complementary techniques in our statistical analysis \citep[as in ][]{Baker2022MNRAS.510.3622B}.
The first of these is Partial Correlation Coefficients analysis. This consists in determining the partial correlation between two quantities whilst holding other quantities constant and enables us to identify the  `true', intrinsic correlations between two quantities and distinguish them from correlations that are indirect and induced by other scaling relations. It is a way of separating out which quantities are truly correlated and which appear correlated as a consequence of dependence on a 3rd  quantity. 
The equation for partial correlation coefficients between two quantities A, and B, whilst controlling for C, is given by 
\begin{equation}
    \rho_{AB|C}=\frac{\rho_{AB}-\rho_{AC}\rho_{BC}}{\sqrt{1-\rho_{AC}^2}\sqrt{1-\rho_{BC}^2}},
\end{equation}
where $\rho_{XY}$ is the Spearmann rank correlation coefficient between quantities $X$ and $Y$.
The Partial Correlation Coefficients can be defined also in the case of multiple variables. In this case they provide the partial correlation between two quantities while controlling for all the others.
An important limitation of the Partial Correlation Coefficients is they are fully meaningful only for {\it monotonic} relations.

When exploring the dependence of a quantity A on two other quantities, B and C, it is often very useful to visualize the dependence by showing the distribution of points on of B (y-axis) versus C (x-axis) plot, where the points are colour-coded by the A (z-axis) quantity.
The partial correlation coefficients are an effective method to derive the `arrow' pointing towards the steepest average gradient of increasing A, hence allowing the determination of the relative role of the two quantities B and C in driving A.
The arrow angle is measured from the horizontal (from the 3 o' clock position). The equation for the arrow angle is \citep{2020BluckB}
\begin{equation}
    \text{tan}(\theta)=\frac{\rho_{BA|C}}{\rho_{CA|B}}
    \label{theta}
\end{equation}
where $\theta$ gives the arrow angle and $C$ is the $x$ axis quantity, $B$ is the $y$ axis quantity and $A$ the $z$ axis quantity. 

The second technique is the Random Forest regression, which is a form of supervised (meaning identifiable labels) machine learning. Random Forest regression is a way of using multiple decision trees to identify the most important parameters for accurately determining a certain target. The data is split into a test and train set. In the training set, the target quantity is removed leaving the features that could contribute to it. The random forest algorithm builds a model based upon this training set by attempting to decrease Gini Impurity (a measure of the quality of a split in the decision tree). This model is then applied to the test sample (which it has not seen before) in order to find the parameter importances of the features in determining the target. In order to check that the method is not overfitting the data, i.e. that is not fitting the noise of the training data, comparison is made between the mean squared error (MSE) of the training sample and the test sample. If the MSE of the training sample is much less than the test sample, it means that the model overfits. If the MSE of the test sample is less than the training sample, it signifies a case of underfitting. We tune the hyper-parameters (such as the number of trees in the forest and the minimum number of samples per leaf) in order to minimise both overfitting and underfitting. One of the benefits of Random Forest regression is that it can uncover highly non-linear and even non-monotonic trends and so is unconstrained by the monotonicity requirement of partial correlation coefficients. It can also investigate all the potential parameters simultaneously and can distinguish between indirect and intrinsic dependencies \citep[for an in-depth analysis and discussion of the random forest regression method see][]{2020Bluck,2020BluckB, Bluck2022A&A...659A.160B}.

\section{Local metallicity dependence on global and local quantities}

\subsection{Resolved FMR in MaNGA}

Figure \ref{fig:resolved_FMR_manga} shows the star formation rate surface density ($\Sigma _{SFR}$) as a function of the stellar mass surface density ($\Sigma_*$), colour coded by the local gas-phase metallicity (12+log(O/H)) for all spaxels in the MaNGA sample, selected as discussed above.  The data is binned in an hexagonal grid, where the colour corresponds to the mean metallicity of the spaxels in each hexagonal bin.  Each bin contains at least 150 spaxels (this is the same for all figures of this type involving the resolved quantities in the MaNGA sample).
The contours give the density distribution of spaxels, where the outer density contour encloses 90\% of the spaxels. The colour shading provides an indication as to how much $\Sigma_{SFR}$ and $\Sigma_*$ drive the metallicity. If it was purely $\Sigma_*$ we would expect vertical bands of colour, i.e. at fixed $\Sigma_*$ we would expect no variation in metallicity with varying $\Sigma_{SFR}$, whilst if it was just $\Sigma_{SFR}$ we would expect horizontal bands of colour.
The colour shading clearly indicates that the local metallicity correlates, {\it separately}, {\it both} with
$\Sigma _*$ {\it and} (inversely) with $\Sigma _{SFR}$
, i.e. it shows the existence of a spatially resolved, local FMR: at a fixed $\Sigma _{SFR}$ the metallicity depends on $\Sigma _*$ and, viceversa, at a fixed $\Sigma _*$ the metallicity depends (inversely) on $\Sigma _{SFR}$. 

The inclination of the colour shading clearly indicates that the local metallicity depends primarily on $\Sigma _*$ and, to a lower extent, on $\Sigma _{SFR}$.
The relative roles of $\Sigma _*$ and $\Sigma_{SFR}$ in driving the local metallicity can be quantified through the
 Partial Correlation Coefficient arrow, as defined in the previous section, which is drawn in Figure \ref{fig:resolved_FMR_manga} and its angle relative to the horizontal is given in the top-left corner (its error is obtained by bootstrap random sampling 100 times).
The arrow angle (and direction) show that, in order to increase the metallicity, the most important parameter to increase is $\Sigma_*$, as would be expected by the resolved MZR, but that this relation has a secondary dependence on $\Sigma _{SFR}$, where the metallicity increases as $\Sigma _{SFR}$ decreases (at fixed $\Sigma_*$). The arrow inclination relative to the horizontal (which would give pure dependence on 
$\Sigma_*$) is 40$^\circ$, i.e. on average $\Sigma_*$ contributes 56\% to the metallicity variation, while $\Sigma_{SFR}$ contributes for 44\%. This ratio is close to what found for the global FMR by \cite{2010Mannucci}.

Figure \ref{fig:FMR_tracks} shows the metallicity as a function of the stellar mass surface density binned by $\Sigma _{SFR}$ and is directly comparable to the FMR representations in \citet{2010Mannucci}.
This figure provides essentially the same information as  Fig.\ref{fig:resolved_FMR_manga}: the different lines in Figure \ref{fig:FMR_tracks} show cuts of the 3D surface in horizontal lines of constant $\Sigma _{SFR}$.
The value in each track is given by the mean $\Sigma_*$ and $\Sigma _{SFR}$ in that bin (we found no significant difference when using the median). 
The shaded region gives the standard error of the mean of $\Sigma _{SFR}$ in each bin. The black dashed line shows the resolved mass-metallicity relation without binning in  $\Sigma _{SFR}$.

Figure \ref{fig:FMR_tracks} shows a resolved FMR in MaNGA (further supporting what already seen in Figure \ref{fig:resolved_FMR_manga}): while there is a clear correlation between metallicity and $\Sigma _*$, at a given $\Sigma _*$ the metallicity anti-correlates with $\Sigma _{SFR}$. Yet, at high $\Sigma _*$ and low $\Sigma _{SFR}$ the correlation is inverted. This inversion was seen also in the integrated FMR \citep[e.g.][]{Yates12,Maiolino2019,Trayford_and_Joop_2019MNRAS.485.5715T,2020CurtiMNRAS.491..944C,Kumari21}. The regime at high stellar masses and low SFR, and specifically low $\Sigma _{SFR}$ is where the contribution of DIG might contaminate the nebular lines \citep{Belfiore22,Tacchella22}; it is also the regime where the nebular lines are faintest and, therefore, where biases in the metallicity determination or S/N cuts may play a role \citep{Salim2014ApJ...797..126S, 2020CurtiMNRAS.491..944C}. 
Although we have tried to minimise these potential issues through our stringent spaxel selection criteria, they may still play a role in explaining the observed FMR inversion. It remains unclear whether this inversion is a real effect or a product of residual biases (S/N cuts) or the calibrations. 
Another potential explanation could be that quiescent regions (typically depleted of gas) are selectively seen in nebular lines only when they accrete some (low-metallicity) gas: in these regions star formation may remain inefficient \citep{Piotrowska2019}, hence not contributing to a star formation enhancement, but the accreted gas would dilute the metallicity \citep{Kumari21}. Another possibility, suggested by recent simulations, is that the inversion of the rFMR at high masses and high $\Sigma *$ is due to the feedback effect of AGN \citep{Trayford_and_Joop_2019MNRAS.485.5715T, De_Rossi_2017MNRAS.472.3354D}.

Next we  investigate whether the shape of the resolved FMR depends on the total stellar mass $M_*$ by replotting the $\Sigma _* - \Sigma _{SFR}$ relation color-coded by metallicity, as in Fig.\ref{fig:resolved_FMR_manga}, in three bins of {\it total} stellar mass. These are shown in Figure \ref{fig:multiplot}, where in each of the three panels the contours provide (as reference) the same total distribution of all galaxies. Clearly, the metallicity gradient is different for the three total-mass bins, with the secondary dependence on $\Sigma _{SFR}$ being most prominent in low mass galaxies and least important in high mass galaxies. This effect is also quantified by the Partial Correlation arrows, shown in the three panels. This trend is similar to what seen in the global FMR, whereby the secondary dependence on {\it total} SFR is weaker, or even vanishes, in massive galaxies \citep{2010Mannucci,2020CurtiMNRAS.491..944C}. In addition, we re-plot Figure \ref{fig:FMR_tracks} in three mass bins as shown in Figure \ref{fig:multiplot_tracks}. We can see that as the total stellar mass increases the separation between the separate tracks decreases (signifying the weakening effect of $\Sigma _{SFR}$ on the rFMR at high stellar masses). Summarising, these plots indicate that the local metallicity depends {\it also} on the global stellar mass, M$_*$, in addition to the dependence on $\Sigma _*$ and $\Sigma _{SFR}$.
The dependence of the rMZR on {\it total} stellar mass has been highlighted previously in \citet{Barrera_global_from_local_radial2016MNRAS.463.2513B}, \cite{Hwang2019ApJ...872..144H}, \citet{Gao2018ApJ...868...89G} and \cite{Boardman22}, but these works did not addess the relative importance of both local and integrated mass and SFR. rMZR dependence on total stellar mass has also been found in simulations \citep{Trayford_and_Joop_2019MNRAS.485.5715T}.

The next sections will explore and disentangle these dependencies in more detail.

\subsection{Parametrisation of the resolved FMR}

\label{sec:param}

\begin{figure*}
    \centering
    \includegraphics[width=2\columnwidth]{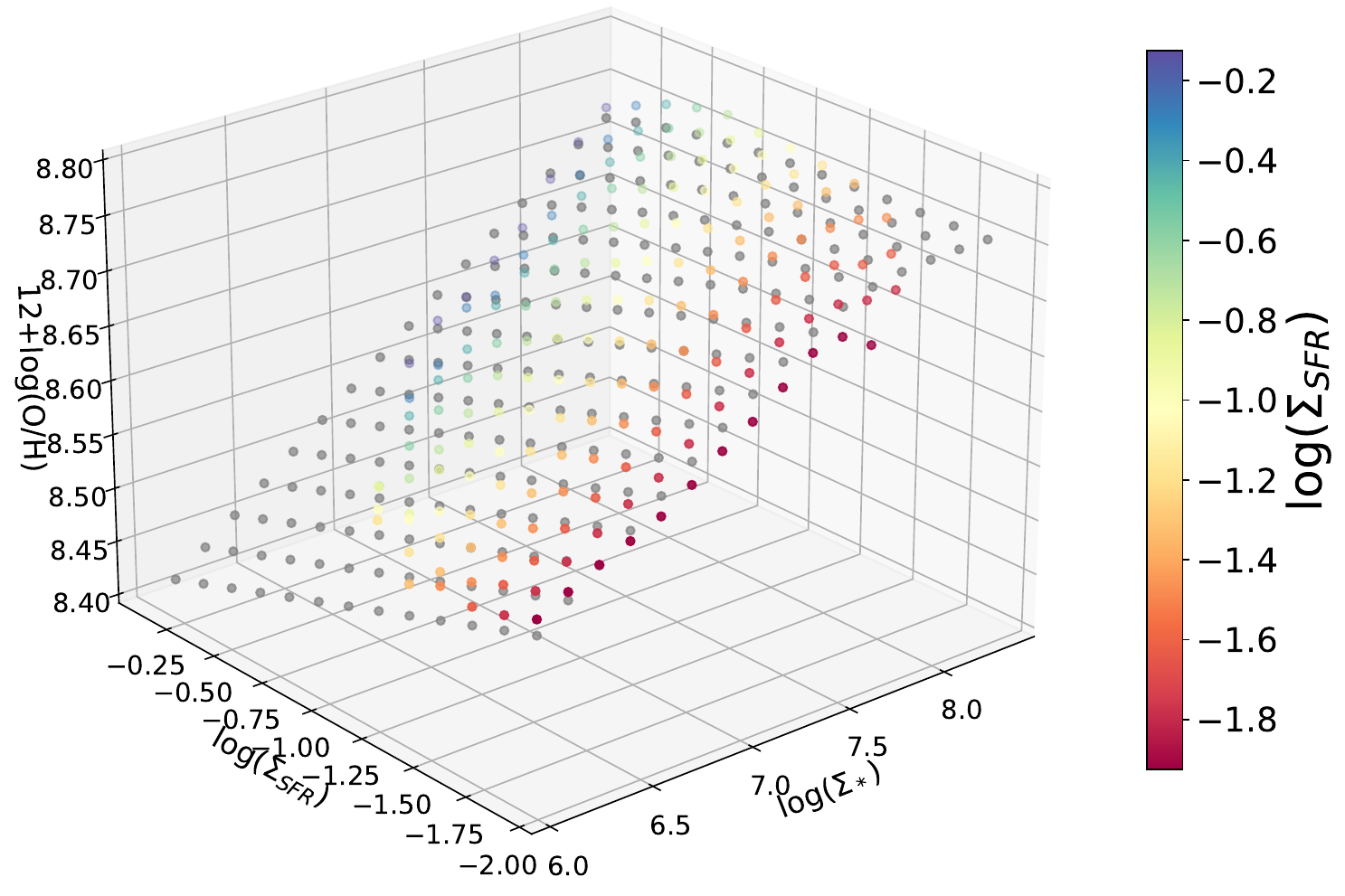}
    \includegraphics[width=1\columnwidth]{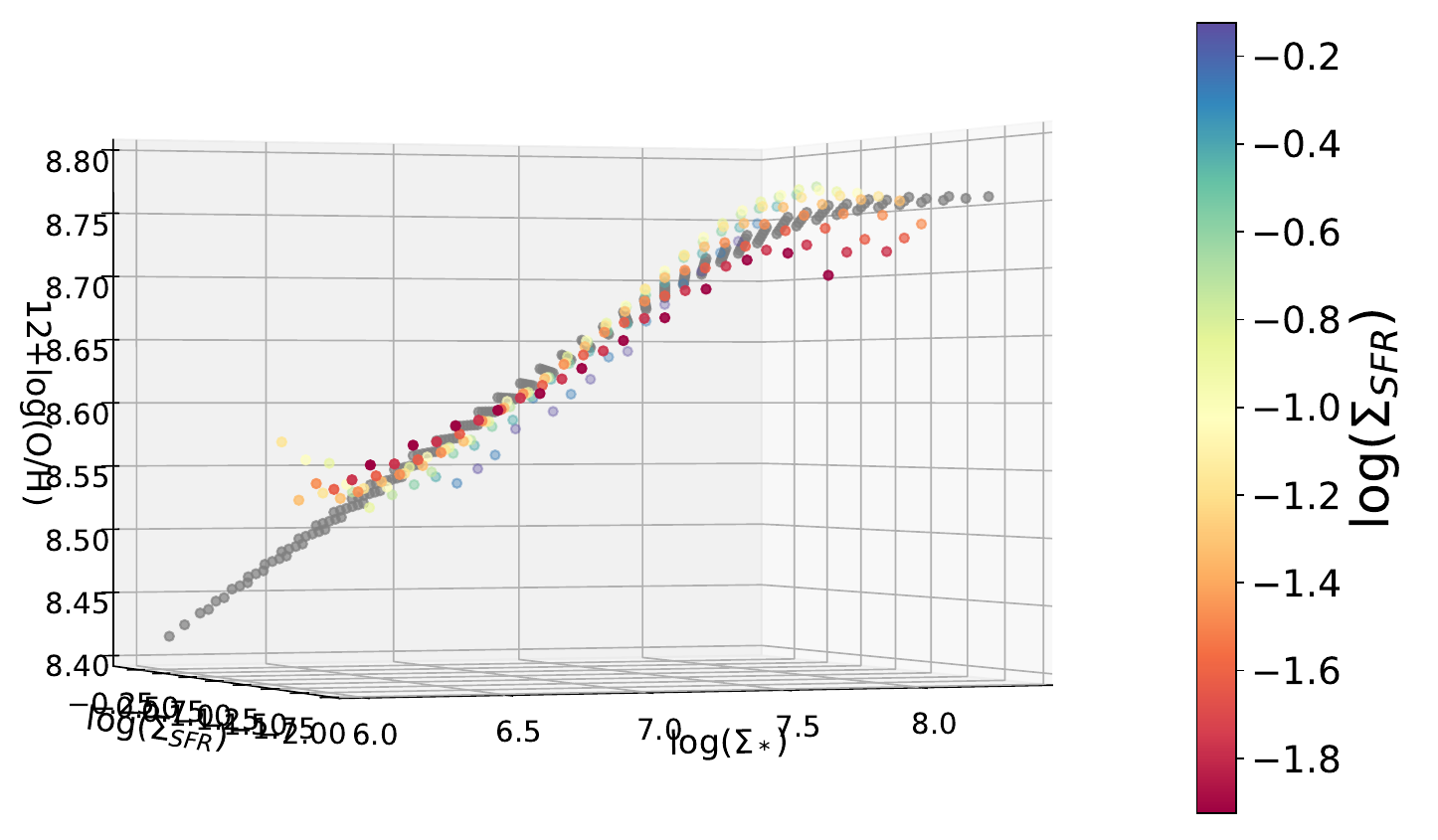}
    \includegraphics[width=\columnwidth]{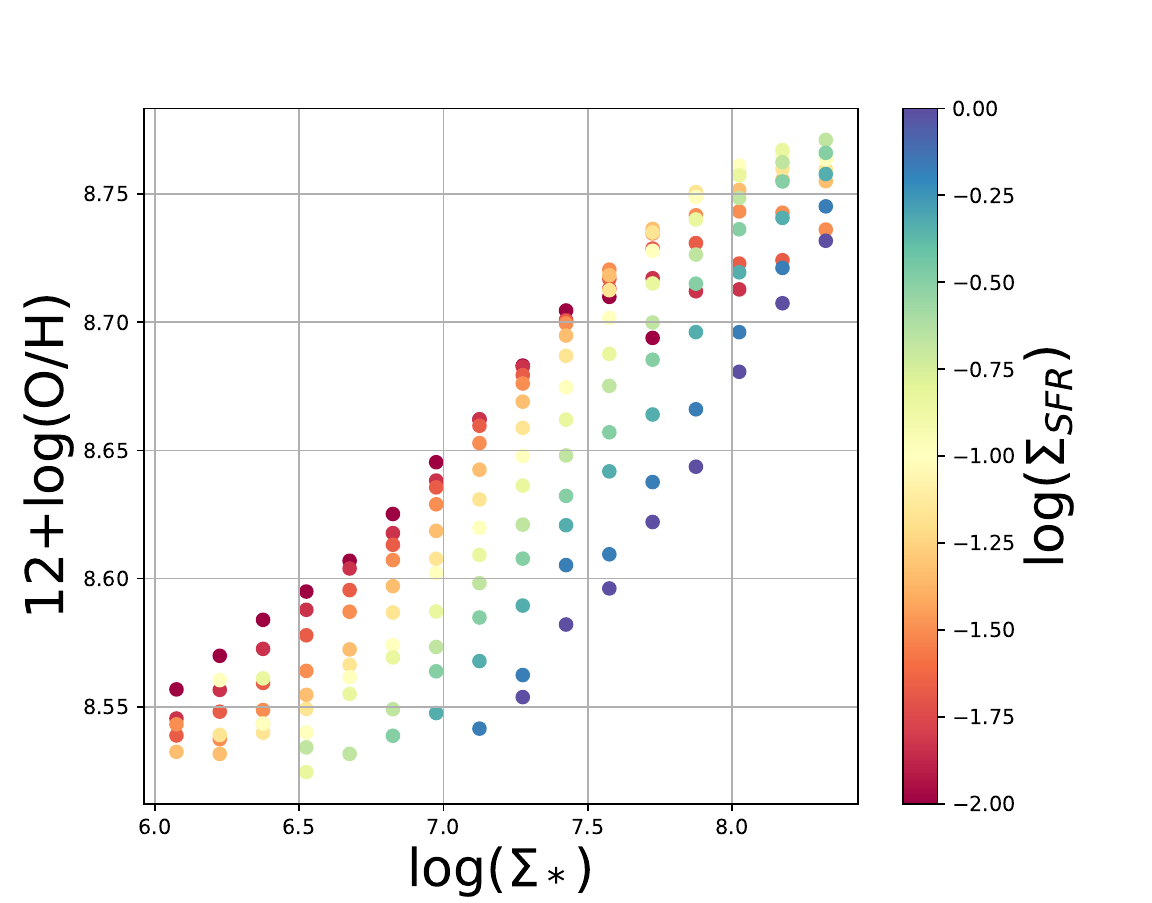}
    \caption{Three varying projections of the resolved FMR, i.e. the relation between metallicity, $\Sigma_*$, $\Sigma_{\rm SFR}$. Coloured circles indicate the mean values of the metallicity in  bins of $\Sigma_*$ - $\Sigma_{\rm SFR}$ whilst the grey circles are the 2D surface fit to the metallicity for the parameterisation given in the text. The colour-coding corresponds to the value of $\Sigma_{\rm SFR}$.
    The lower left panel shows a representation of the rFMR that appears to reduce the scatter in metallicity. The lower right plot shows the metallicity-$\Sigma _*$ 2D projection colour-coded by $\Sigma _{SFR}$, i.e. a different representation of Figure \ref{fig:FMR_tracks}.}
    \label{fig:FMR_param}
\end{figure*}
\citet{2010Mannucci} parameterise the FMR by fitting a 2D surface to M$_*$-SFR space in bins of metallicity. 
 
We follow this approach for the spatially resolved MaNGA data in order to parameterise a resolved FMR surface.
We start by binning spaxels in bins of $\Sigma_*$ and $\Sigma_{\rm SFR}$ and considering the average metallicity in these bins. The use of median values instead does not change our results.

We parameterise the surface using the functional form
\begin{equation}
    12 + \rm log(O/H)=Z_0 - \gamma/\phi\, \rm log\bigg(1+\bigg(\frac{\Sigma_*}{M_0(\Sigma_{SFR})}\bigg)^{-\phi}\bigg)
    \label{eq:mirko}
\end{equation}
introduced by \citet{2020CurtiMNRAS.491..944C}, where $Z_0$ characterises an upper metallicity limit. log($M_0(\Sigma_{SFR}))=m_0+m_1\rm log(\Sigma_{SFR})$ is the turnover mass, with a dependence on $\Sigma_{SFR}$, for which, once this mass is reached, the metallicity approaches $Z_0$. $\gamma$ describes a power law behaviour at low stellar masses, and $\phi$ controls the width of the transition region between low and high mass regimes.

We fit this 2D surface to the binned data (for bins with 150 or more spaxels), to obtain the coefficients of the best fit.

The best fit parameters, both for the total rFMR and for the rFMR in three total stellar mass intervals are given in Table \ref{table:fits}.

Figure \ref{fig:FMR_param} shows three projections of the three-dimensional rFMR. The plots show the mean metallicities for bins of $\Sigma_*$-$\Sigma_{\rm SFR}$ with more than 150 spaxels (colour coded by $\Sigma_{\rm SFR}$) and the metallicity calculated from the best fit model using the parametrisation given by Equation \ref{eq:mirko} (grey circles). The lower left plot shows an orientation of the 3D surface that appears to reduce the scatter in the relationship. The lower right plot shows the rMZR obtained by collapsing the 3D rFMR into 2D along the $\Sigma_{\rm SFR}$ axis (i.e. the same diagram as Figure \ref{fig:FMR_tracks} but with a different representation). These plots show that the data agrees well with the parameterisation of the rFMR given by Equation \ref{eq:mirko}, although there are some deviations at the boundaries of the distribution. Extrapolation of the surface beyond the parameters space probed by our MaNGA data is therefore risky.

\begin{figure}
    \centering
    \includegraphics[width=\columnwidth]{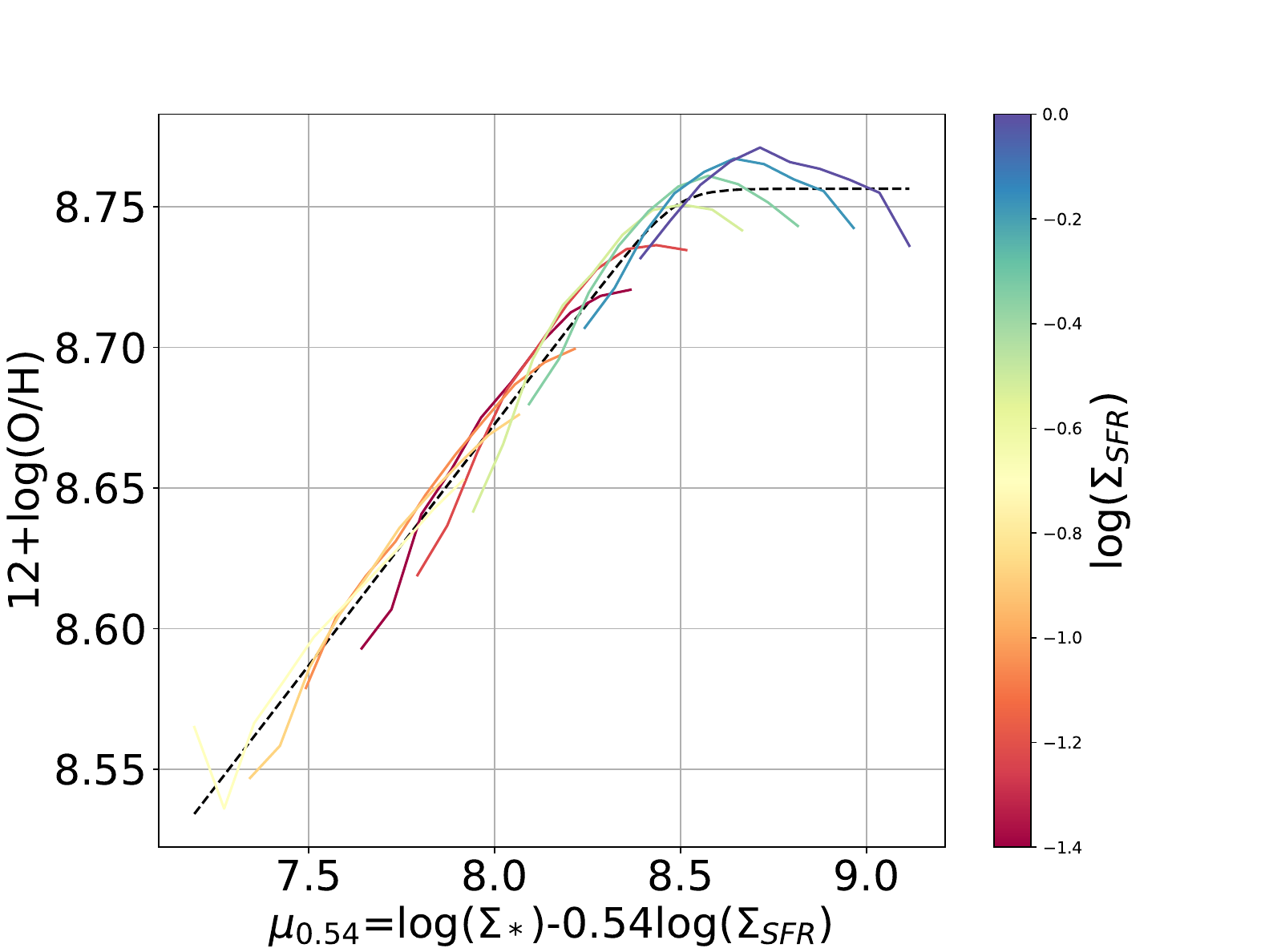}
    \caption{Metallicity versus $\rm \mu _{0.54}= log(\Sigma _*) -0.54log(\Sigma _{SFR})$, i.e. the parameter resulting in a projection of the 3D FMR which gives the minimum scatter in metallicity. Tracks are binned and color-coded in $\Sigma _{SFR}$.
    The black dashed line is the best fit to this projection. Can see how the tracks show minimal dispersion for this projection of the rFMR.}
    \label{fig:best_alpha}
\end{figure}

As can be seen in the lower left plot of Figure \ref{fig:FMR_param}, it is possible to find a projection of this 3D plot that reduces the scatter. In order to find the optimal projection we follow the approach of \citet{2010Mannucci} by defining the quantity $\mu_\alpha$ as a linear combination of $\Sigma_*$ and $\Sigma_{\rm SFR}$
\begin{equation}
 \rm   \mu_\alpha=log(\Sigma_*) -\, \alpha\,log(\Sigma_{\rm SFR}).
 \label{eq:mu_alpha}
\end{equation}
Equation \ref{eq:mu_alpha} describes a rotation of the $\Sigma_*$ and $\Sigma_{\rm SFR}$ axes. We then calculate the metallicities in bins of $\mu_\alpha$ and find the value of the parameter $\alpha$ that minimizes scatter in the
 $\mu$-metallicity relationship.
 In Appendix C we show Figure \ref{fig:alpha} which shows the mean dispersion of the metallicity as a function of $\alpha$. The minimum dispersion is given by $\alpha =0.54$.  
 
When performing this test for the global properties, \citet{2010Mannucci} found a value of 0.32 for the equivalent parameter $\alpha$, but in their revisited work \citet{2020CurtiMNRAS.491..944C} found a value of 0.55, hence very similar to that found in this work, indicating that the best projection for the rFMR is the same as for the integrated scaling relation.

Figure \ref{fig:best_alpha} shows the  metallicity--$\mu_{0.54}$ relation for the individual $\Sigma_{\rm SFR}$ tracks (colour-coded by $\Sigma_{\rm SFR}$ as before) enabling us to see how the projection reduces the scatter relative to the un-projected plot (Fig. \ref{fig:FMR_tracks}). 
The black line corresponds to the best fit to this projection.
and is given by \citep[][]{2020CurtiMNRAS.491..944C}
\begin{equation}
    12 + \rm log(O/H)=\rho - \frac{\zeta}{\psi}\rm log\bigg(1+\bigg(\frac{\mu_{0.54}}{\omega}\bigg)^{-\psi}\bigg)
\end{equation}
where $\rho$=8.756, $\zeta$=0.172, $\psi$=8.542, and $\omega$=8.487.

The data (coloured tracks) closely follow the best fit with little scatter. 

 We investigate the effect of integrated mass on the rFMR by splitting the sample into three separate bins of stellar mass as in the previous section. We then repeat the previous steps and minimise the metallicity scatter in each of the three mass bins, resulting in the best-fit parameters reported in Table \ref{table:fits}. The parameter $\alpha$ varies significantly with total stellar mass. We can (crudely) parameterise a $M_*$-$\alpha$ relation with a linear relation of the following form
\begin{equation}
   \rm \alpha=-0.133\,log(M_*/M_\odot)+1.914.
\end{equation}
This relation shows that $\alpha$ decreases with increasing stellar mass, and that therefore the contribution of $\Sigma_{SFR}$ becomes less important. This is in agreement with the qualitative trend seen in Figure \ref{fig:multiplot} where we observed that the relative importance of $\Sigma_{SFR}$ decreases for the larger total stellar mass bins.

\begin{table*}
\caption{Parameters $Z_0$, $\gamma$, $\phi$, $m_0$ and $m_1$ that best fit the rFMR surface by using the analytical Equation  \ref{eq:mirko}, for the full MaNGA sample and for the three mass bins given in the first column. The seventh column gives the parameter $\alpha$ that minimises the metallicity dispersion when the latter is expressed as a function of $\mu _\alpha = \Sigma _*-\alpha \Sigma _{SFR}$. The last column gives the resulting minimum metallicity dispersion.}
\centering
\label{table:fits}
\begin{tabular}{l c c c c c c c}
\toprule
\multirow{2}{*}{Mass bin} & \multirow{2}{*}{$Z_0$} & \multirow{2}{*}{$\gamma$} & \multirow{2}{*}{$\phi$} & \multirow{2}{*}{$m_0$} & \multirow{2}{*}{$m_1$} & \multirow{2}{*}{$\alpha$} & \multirow{2}{*}{$\sigma$}\\[+8pt]
\midrule

9.0<$M_*$<9.8 & 8.71 ($\pm$ 0.03) & 0.13 ($\pm$ 0.3) & 2.66 ($\pm$ 0.6) & 8.88 ($\pm$ 1.0) & 0.76 ($\pm$ 0.06) & 0.67 & 0.064 \\[+10pt]

9.8<$M_*$<10.6 & 8.75 ($\pm$ 0.03) & 0.097 ($\pm$ 0.3) & 2.40 ($\pm$ 0.7) & 8.44 ($\pm$ 1.2) & 0.31 ($\pm$ 0.08) & 0.57 & 0.050 \\[+10pt]

10.6<$M_*$<11.4 & 8.78 ($\pm$ 0.18) & 0.097 ($\pm$ 0.3) & 2.02 ($\pm$ 0.7) & 8.51 ($\pm$ 2.8) & 0.28 ($\pm$ 0.12) & 0.46 & 0.044\\[+10pt]

 Combined & 8.75 ($\pm$ 0.015) & 0.14 ($\pm$ 0.07) & 2.73 ($\pm$ 0.4) & 8.58 ($\pm$ 0.5) & 0.50 ($\pm$ 0.04) & 0.54 & 0.060\\[+10pt]

\bottomrule
\end{tabular}
\end{table*}

\subsection{The FMR in MaNGA and Legacy SDSS}

\begin{figure}
    \centering
    \includegraphics[width=\columnwidth]{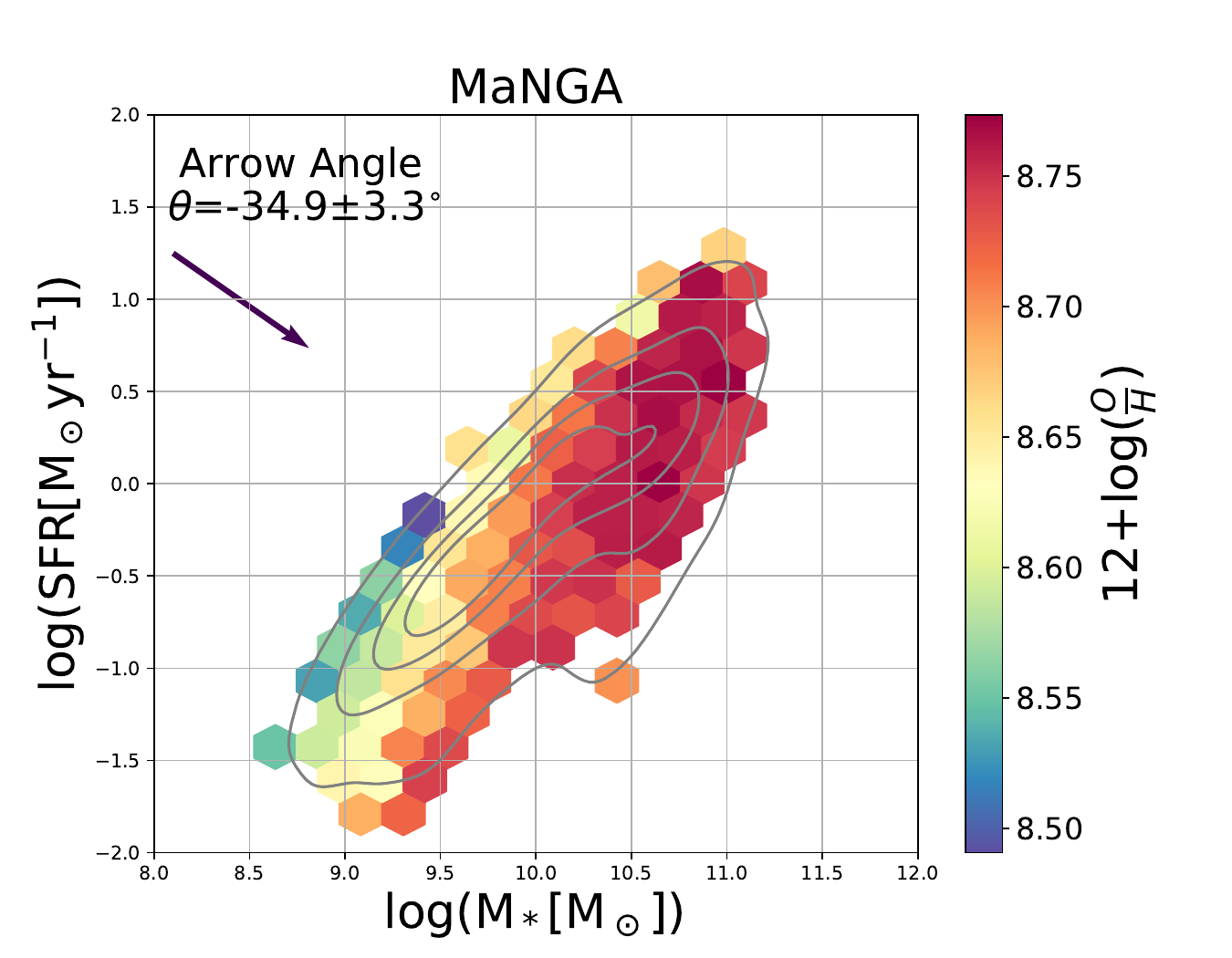}
    \includegraphics[width=\columnwidth]{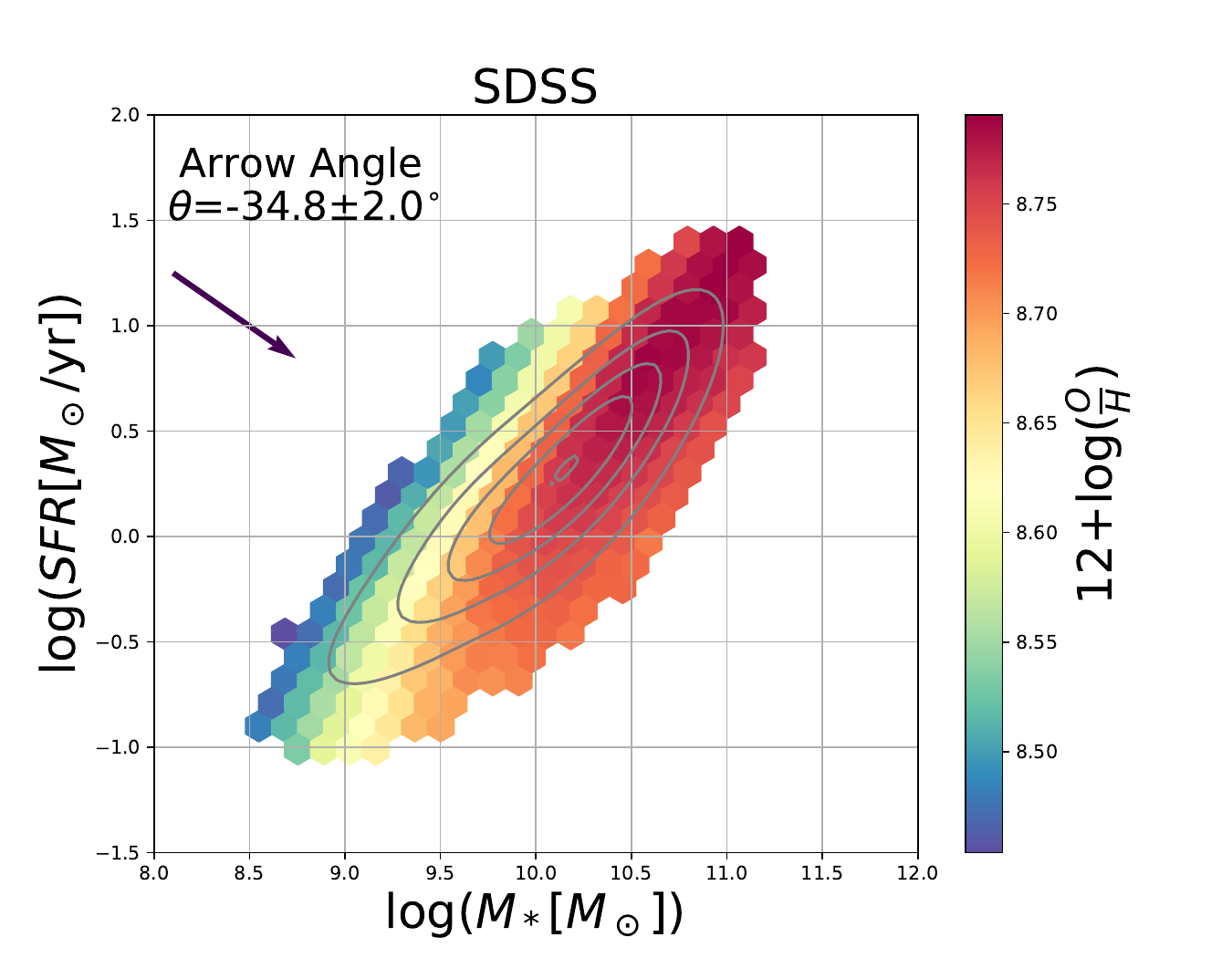}
    \caption{
    Global star formation rate (SFR) versus global stellar mass (M$_*$) for galaxies, colour coded by the global metallicity. Top: MaNGA sample; bottom: SDSS sample. Each hexagon shows the mean metallicity of galaxies in that bin. The arrow angle gives direction of the steepest gradient, averaged across the entire distribution, as obtained by the Partial Correlation Coefficients (see Sec. \ref{sec:methodology}) and is measured from the horizontal. The figure shows that the galaxies in the subset of MaNGA galaxies reassuringly follow the same global FMR as the SDSS parent sample, although with lower (hence noisier) statistics. The density contours give the distribution of the galaxies in each diagram, where the outer density contour encloses 90\%.
    }
    \label{fig:global_fmr}
\end{figure}

\label{sec:SDSSvMaNGA}

Before exploring the simultaneous dependence on local and global properties,
we investigate the existence and shape of the global FMR in the MaNGA sample and compare it with the FMR seen in the larger Legacy SDSS survey \citep[][]{SDSS_DR7_2009ApJS..182..543A}. We have re-derived the global FMR for the legacy SDSS from the MPA-JHU catalog \citep[][]{2004MNRASBrinchmann} by using selection criteria consistent with the one adopted for the MaNGA analysis discussed above (S/N on the nebular lines, BPT selection, maximum inclination cut), resulting in a total of $\sim$179100 galaxies. Global stellar masses and star formation rates are taken from the MPA-JHU catalog\footnote{https://wwwmpa.mpa-garching.mpg.de/SDSS/DR7/} \citep{2004MNRASBrinchmann}. Figure \ref{fig:global_fmr} shows the global SFR versus global stellar mass colour coded by the mean metallicity (the average metallicity of star-forming spaxels within 1$r_e$) for the MaNGA sample (upper panel) and for the SDSS sample (lower panel). 
The MANGA diagram is more noisy than the SDSS as a consequence of the lower statistics in terms of the number of galaxies (2002 in MaNGA vs 179100 in SDSS DR7), however the global FMR is clearly seen also for the MaNGA sample. Moreover, as indicated by the Partial Correlation Coefficient arrows, the two global FMRs are fully consistent with each other, in terms of relative dependence of the metallicity on mass and SFR (same inclination of the PPC arrow). This confirms that the MaNGA sample is not biased for what concerns the FMR.

We note here that a test of the effect of different aperture sizes between the two surveys is carried out in Appendix \ref{sec:app_test} and aperture effects are found to be unimportant here (we find the physical  aperture for our MaNGA sample is 3kpc, for SDSS 3.1kpc). In addition, we note the values of the star formation rates for the MaNGA galaxies (obtained from PIPE3D) are lower than those for Sloan (obtained from the MPA-JHU catalogue) despite using the same initial mass functions - this is a known difference between the two catalogues (they use different conversions) and we again reiterate that our focus here is the comparison between the arrow angles for the two surveys, i.e. that the two sample have the same global FMR dependences.

\begin{figure}
    \centering
    \includegraphics[width=\columnwidth]{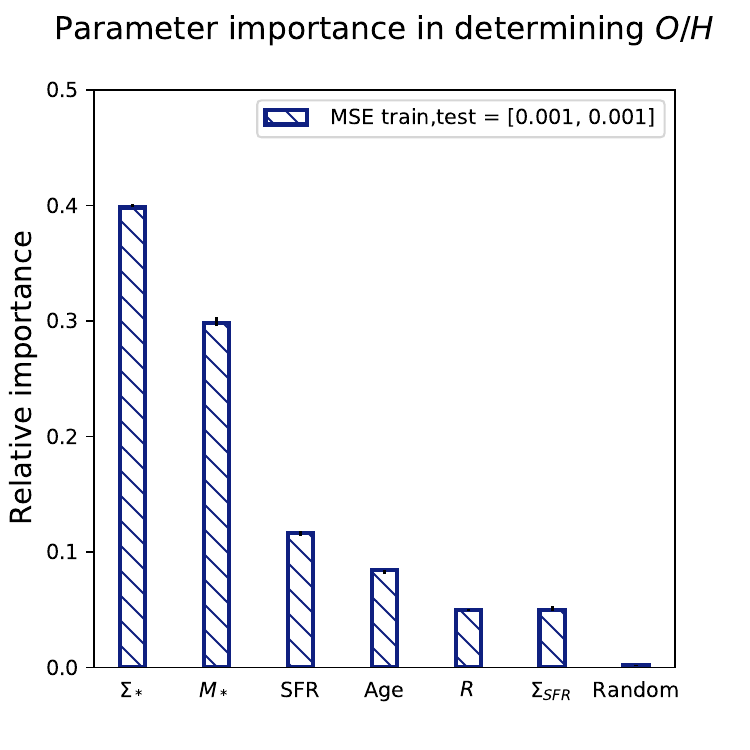}
    \caption{Importance of various global and local parameters for determining the resolved metallicity, based on the Random Forest (RF) regression. The parameters included are:  resolved stellar mass ($\Sigma_*$), total stellar mass M$_*$, total star formation rate (SFR),  mass-weighted age of the stellar population, galactocentric radius (R, in units of $R_e$),  resolved star formation rate ($\Sigma_{SFR}$), and a (control) uniform random variable. The RF clearly indicates that {\it both} resolved and global stellar mass play the most important role in determining the local metallicity. The SFR, both global and resolved, plays a secondary role. Galactocentric distance also seems to have a (smaller) direct impact on the local metallicity.  The MSE for the test and train samples are included to ensure no presence of overfitting or underfitting.}
    \label{fig:RF_Z}
\end{figure}

\begin{figure}
    \centering
    \includegraphics[width=\columnwidth]{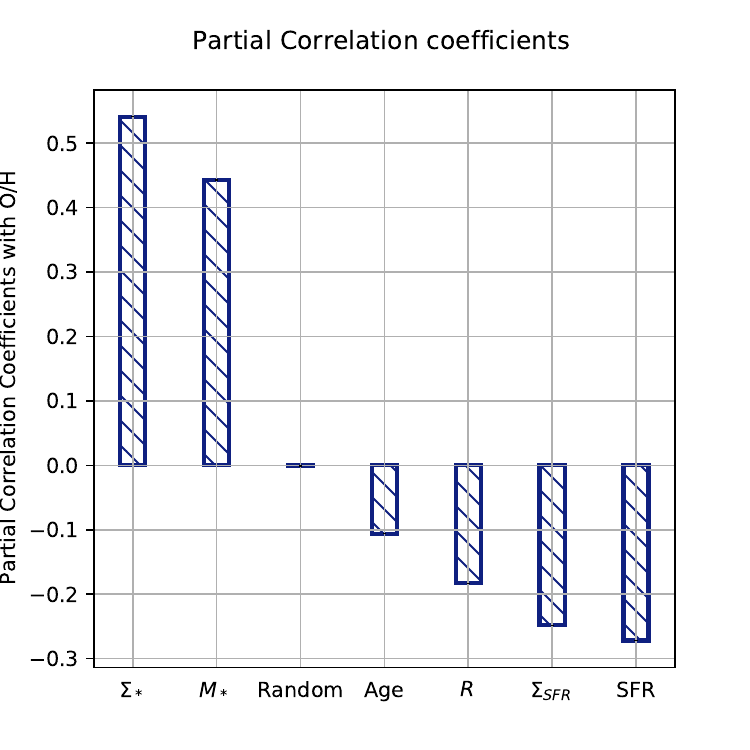}
    \caption{Partial Correlation Coefficients (PCC) between metallicity and, the resolved stellar mass $\Sigma_*$, total stellar mass M$_*$, a uniform random variable (Random), the mass-weighted age of the stellar population, the total star formation rate (SFR), the resolved star formation rate ($\Sigma_{SFR}$), and the galactocentric radius (R, in units of $R_e$). As for the RF, the PCC clearly indicate that {\it both} resolved and global stellar mass play the most important role in determining the local metallicity. Both resolved and global SFR play a (secondary, inverse) role in determining the local metallicity. The PCC also highlight more strongly a direct role of the galactocentric distance in determining the local metallicity.}
    \label{fig:pcc_manga}
\end{figure}

\begin{figure}
    \centering
    \includegraphics[width=\columnwidth]{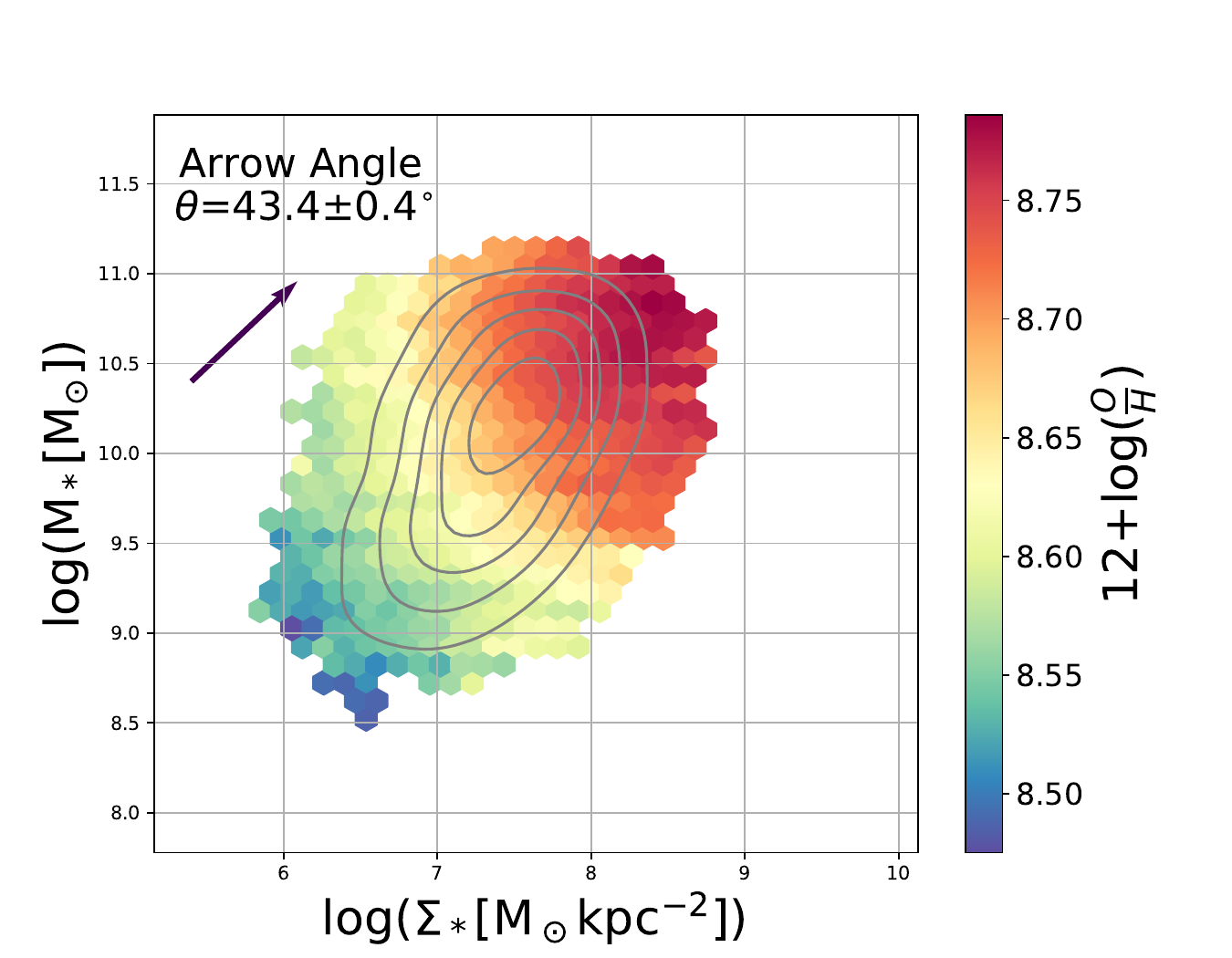}
    \caption{Total stellar mass $M_*$ versus stellar mass surface density $\Sigma _*$, colour coded by local metallicity. The arrow gives the direction of steepest gradient across the whole distribution as inferred from the Partial Correlation Coefficients. The figure illustrates more clearly that the local metallicity depends {\it both} on local and global stellar mass. Contours show the density distribution of spaxels.}
    \label{fig:res_v_glob}
\end{figure}

\subsection{Simultaneously exploring the metallicity dependence on resolved and global properties}
\label{sec: res and glob}
In this section we explore the relative roles of local and global physical properties in driving local metallicity. Answering this question helps us to understand whether global metallicity scaling relations simply stem from the resolved counterparts or if global properties have a direct role in determining the metallicity. As a consequence, this kind of analysis provides information on the physical processes at work in galaxy evolution.

\subsubsection{Simultaneous analysis of all quantities}

Figure \ref{fig:RF_Z} shows the importances, obtained from the RF analysis, for determining the local metallicity, of the following local and global parameters:  $\Sigma_*$, $M_*$, $\Sigma _{SFR}$, SFR, galactocentric radius (R, measured in units of the effective radius), the mass-weighted age of the stellar population (Age), and a uniform random variable, which serves as a control variable to verify that a random quantity has no predictive power. The errors have been obtained by bootstrap random sampling 100 times, with the sample size equal to the number of spaxels (Note that error bars are barely visible in the Figure). 
The RF bar-chart shows that the two most important parameters for determining metallicity are the resolved and global stellar mass. 

The same result is obtained by computing the Partial Correlation Coefficients  between the local metallicity and other global and resolved quantities, while controlling for $\Sigma_*$ and $M_*$. The results are shown by the bar chart in Figure \ref{fig:pcc_manga}. Errors are obtained by bootstrap random sampling and are set as error-bars in the figure, but are too small to be visible. In contrast to the RF, the partial correlation coefficients also provide the sign of the correlations. Also in this case, $\Sigma _*$ and $M_*$ are the most important parameters in driving the local metallicity. We note that there are differences between the random forest regression parameter importances and the partial correlation coefficients (i.e. Age is more important than R or $\Sigma_{SFR}$ in the random forest but less important in the PCC analysis). The differences are likely due to the fact that the Random Forest can also cope with non-monotonic trends, while the partial correlation coefficients are suited only for monotonic relations. Therefore, the slight discrepancies may indicate, in the multi-dimensional space, some of the correlations have some non-monotonicity. Anyway, what both these methods show us is that there appears to be intrinsic dependencies on all of these secondary quantities (but that they are significantly less important than $\Sigma_*$ and $M_*$).
We use the random forest as our baseline for the relative importances, and the PCCs primarily to inform us of the direction of the relationship. 

\subsubsection{The primary metallicity dependence on both local and global stellar mass}

Both the RF and the PCC analysis indicate that the local metallicity depends primarily on {\it both} local and global stellar mass. This is visualised more directly in Figure \ref{fig:res_v_glob}, which shows the global versus resolved stellar mass colour-coded by local metallicity: at a given global $M_*$ the metallicity depends on $\Sigma _*$ and, viceversa, at a given $\Sigma _*$ the local metallicity depends on $M_*$.
This consistent with the same result obtained by \cite{Boardman22, Gao2018ApJ...868...89G}, however we quantify it by also illustrating the PCC arrow, which has a angle of
 43.4$\degree$, indicating that each of $\Sigma _*$ and $M_*$ play a similarly important role in determining the metallicity, confirming what already inferred from Figs. \ref{fig:RF_Z} and \ref{fig:pcc_manga}.

These findings indicate that the global MZR ($M_*-Z$ relation) does not result from its local, resolved version, the rMZR ($\Sigma _*-Z$ relation), but that there is also a direct, intrinsic dependence of the metallicity on the global stellar mass. The implication is that the local metallicity does not depend only on the local metal production and/or the local potential well of the galactic disc, but also depends on the global metal production/retention.

Based on the RF (Figure \ref{fig:RF_Z}) and PCC (Figures \ref{fig:pcc_manga} and \ref{fig:res_v_glob}) analyses, both $\Sigma _*$ and $M_*$ quantities are important in driving the local metallicity (as also indicated by the PCC arrow in Figure \ref{fig:res_v_glob} which, already mentioned, is oriented close to 45$^\circ$), implying that both global and local stellar mass play a similar role in determining the local metallicity.

\subsubsection{The secondary metallicity dependence on SFR and $\Sigma _{SFR}$}

Regarding the secondary (inverse) correlation with SFR, Figure \ref{fig:RF_Z} and Figure \ref{fig:pcc_manga} indicate that the local metallicity does not (inversely) depend only on $\Sigma _{SFR}$, but also, and even more strongly, on the global SFR.
The presence of a direct relation with the global SFR suggests that the dilution scenario (gas accretion that locally dilutes the metallicity and locally boosts the SFR) cannot fully explain the existence of the FMR as it would be fully dependent on the resolved, local SFR. We are aware that, as we use H$\alpha$ based SFRs, we are tracing star-formation (SF) on short ($\sim 10$Myr) timescales. However, it is this order of timescale we need to be sensitive to so that we can investigate the cause of the rFMR. Longer timescales ($\sim 100$Myr) would be of the order of the galaxy dynamical timescales and, therefore, any local dilution effect would be lost.

\subsubsection{The dependence on age}

Based on the findings in \citet{Duerte-Puertas-Age2022arXiv220501203D} and \citet[][]{Boardman22} of the possible importance of age in contributing to the metallicity, we also include the mass-weighted age of the stellar population at the effective radius (obtained from PIPE3D \citealt[][]{Sanchez2016}) as a parameter in Figure \ref{fig:RF_Z} and \ref{fig:pcc_manga}. We find that the metallicity appears to have a small but non-negligible dependence on age. If we use Dn(4000) \citep[][]{Balogh_Dn4000_1999ApJ...527...54B} as proxy of age (as in some other works), we find a significantly higher importance of it (higher than SFR and third most important after $M_*$). However, we ascribe this larger importance not to the role of age, but to the fact that Dn(4000) also depends on stellar metallicity \citep{Kauffman_stellarmasses_SFRs_dn40002003MNRAS.341...33K}, which is correlated with gas metallicity. Therefore, part of the higher importance associated with Dn(4000) relative to age, in predicting the gas metallicity, is due to the fact that this parameter partly incorporates the metallicity information. It is also worth noting that Dn(4000) more closely resembles a light-weighted age - hence, it is more sensitive to the recent star formation history.

\subsubsection{Intrinsic dependence on galactocentric radius}

Figure \ref{fig:pcc_manga} demonstrated that metallicity appears to have an intrinsic dependence on the galactocentric radius R, i.e. the distance of the spaxel/region from the centre of the galaxy. The strength of this dependence is different between the RF and the PCC.  
The first reported work demonstrating the existence of the rMRZ \citep[][]{RosalesOrtega2012ApJ...756L..31R} prompted speculation that metallicity gradients could be explained as a direct consequence of the rMZR combined with the radial decrease of the $\Sigma _*$ \citep[][]{Barrera_global_from_local_radial2016MNRAS.463.2513B}. However, our finding of a direct dependence of the local metallicity on galactocentric distance, after accounting for its dependence on $\Sigma _*$, indicates that metallicity gradients have also a (secondary) {\it intrinsic} component. This is also found in \cite{Boardman22}.

\begin{figure*}
    \centering
    \includegraphics[width=\columnwidth]{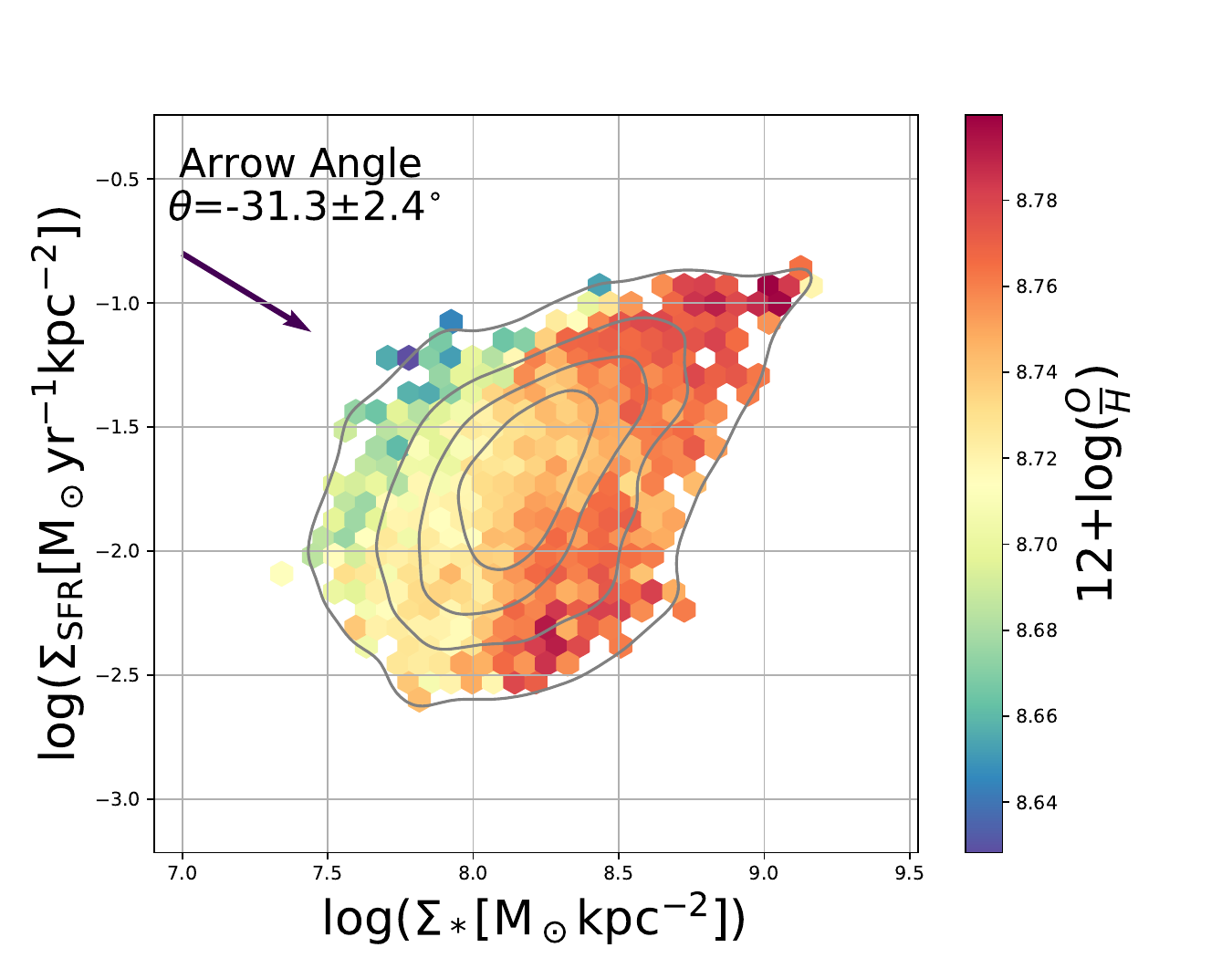}
    \includegraphics[width=\columnwidth]{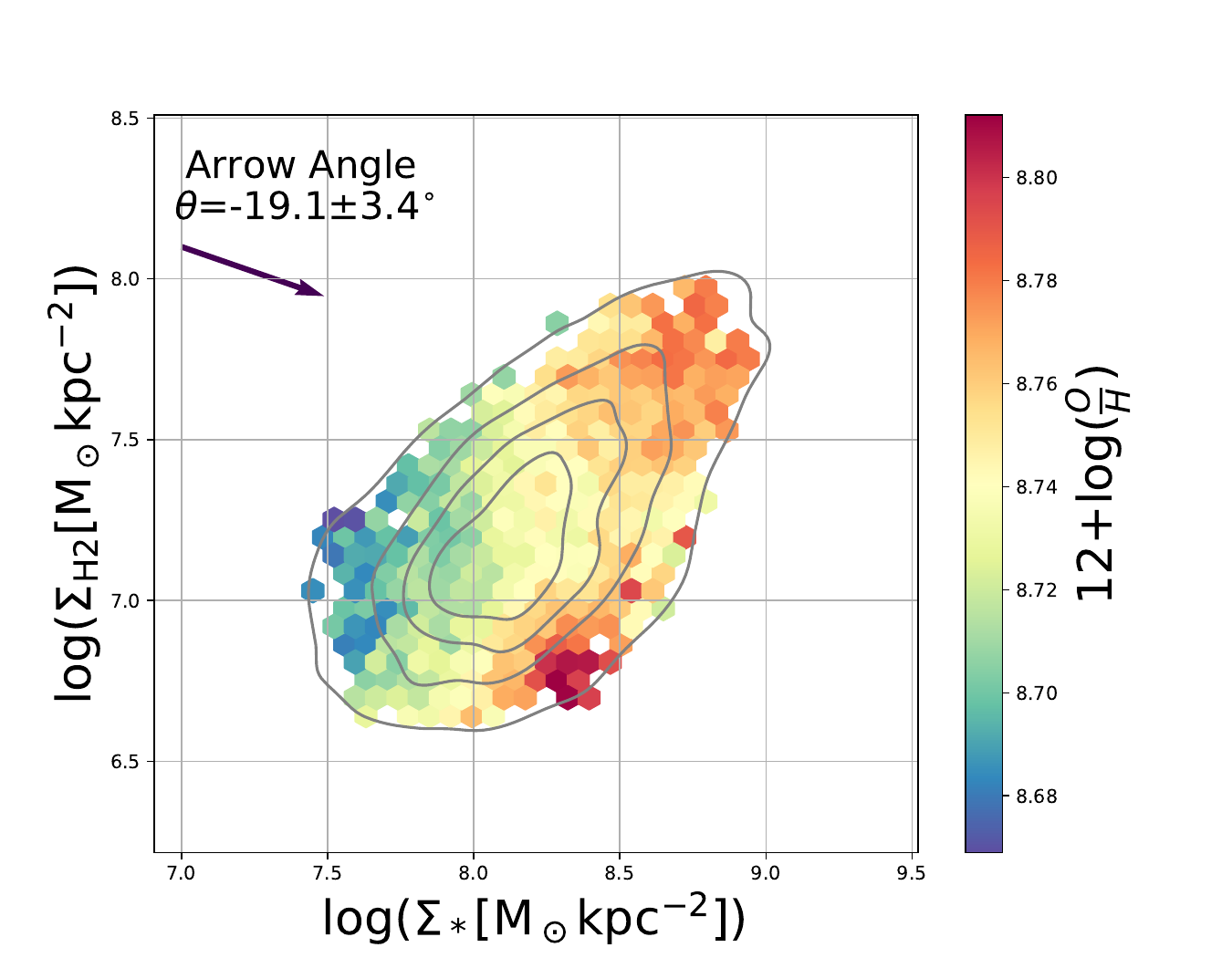}
    \caption{Left: Star formation rate surface density $\Sigma_{SFR}$ versus stellar mass surface density $\Sigma _*$, color-coded by the average local metallicity. Right: Molecular gas mass surface density $\Sigma _{H2}$ versus stellar mass surface density $\Sigma _*$, colour-coded by the average local metallicity. In both panels the contours give the density distribution of spaxels and the arrows give the direction of the steepest gradient across the entire distribution as inferred by the Partial Correlation Coefficients.
    Essentially, the left plot shows the resolved Fundamental Metallicity Relation (rFMR) in the ALMaQUEST sample, i.e. the primary dependence of the metallicity on $\Sigma _*$ along with the secondary inverse dependence on $\Sigma_{SFR}$, while the right plot shows the weaker metallicity dependence on $\Sigma _{H2}$, implying that the rFMR is not a by product of the SFR dependence on gas mass via the S-K relation. }
    \label{fig:M-S-O-Alma}
\end{figure*}

\section{The metallicity dependence on the molecular gas content}

The ALMaQUEST sample enables us to investigate whether metallicity depends on the molecular gas mass surface density ($\Sigma_{H_2}$) and, in particular, whether the secondary dependence on SFR and $\Sigma _{SFR}$ is simply an indirect consequence of a more fundamental dependence on the molecular gas content.

\subsection{The rFMR in ALMaQUEST}

We first verify that the ALMaQUEST subsample (46 galaxies) is not biased in a way that might significantly offset its rFMR from the trend observed in the full MaNGA sample (which after cuts consists of 2002 galaxies).

The left panel of Figure \ref{fig:M-S-O-Alma} shows the resolved main sequence ($\Sigma_{SFR}$ versus $\Sigma_*$) colour-coded by local metallicity. The figure shows clear evidence for a resolved FMR also in the ALMAQUEST sample.
  
The trend is similar to that observed for the full MaNGA sample shown in Figure \ref{fig:resolved_FMR_manga}. More quantitatively, the partial correlation coefficient arrow in the plot (inclination $-31.3\pm 2.4^{\circ}$) is similar to the value obtained in the full MaNGA sample ($-40.1\pm 0.6^{\circ}$). The small (but significant) difference between these arrow angles is because the ALMaQUEST survey contains a larger fraction of massive galaxies than the MaNGA parent sample.
In Appendix \ref{sec:alvman} we show that, for MaNGA, by considering only galaxies within the mass range probed by ALMaQUEST we obtained an FMR consistent with the one present in the ALMaQUEST sample

\subsection{The metallicity dependence on the molecular gas}

\label{sec:alma_rf}

We now explore the dependence of the local metallicity on the resolved molecular gas mass in the ALMaQUEST sample. We do not include any global properties in this analysis as the ALMaQUEST sample is too small for such a study. 

Our baseline is to use the metallicity-dependent $\alpha_{CO}$ conversion factor of \citet{Accurso2017MNRAS.470.4750A}; we will discuss in the appendix the (unrealistic) case of a constant $\alpha_{CO}$.

We start by investigating the metallicity dependence on $\Sigma _{H_2}$ at a given $\Sigma _{*}$.
The right panel of 
Figure \ref{fig:M-S-O-Alma} shows $\Sigma _{H_2}$ versus $\Sigma _{*}$, colour-coded by local metallicity. The PCC arrow confirms a strong metallicity dependence on $\Sigma _{*}$, but a weaker dependence on $\Sigma _{H_2}$ (PCC arrow angle $-19.1\pm3.4\degree$). This is should be directly compared with the left panel of the same figure, where $\Sigma _{H_2}$ is replaced by $\Sigma _{SFR}$ (PCC arrow angle $-31.3\pm2.4\degree$). Clearly, the metallicity dependence on $\Sigma _{SFR}$ is stronger than the dependence on $\Sigma_{H2}$, as quantified by the PCC arrow being more inclined in the former and clearly flatter in the latter.

\begin{figure}
    \centering    \includegraphics[width=\columnwidth]{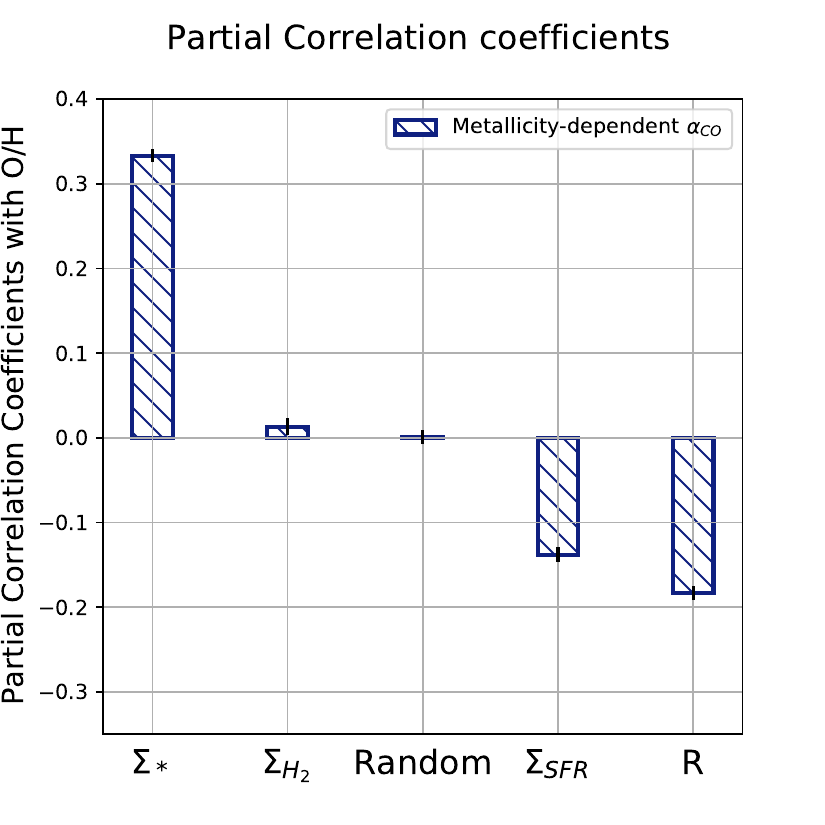}
    \caption{Partial Correlation Coefficients between local metallicity and molecular gas mass surface density ($\Sigma_{H_2}$),  stellar mass surface density ($\Sigma_*$), a uniform random variable (Random),  galactocentric radius (R), and star formation rate surface density ($\Sigma_{SFR}$).  The PCC indicate again that $\Sigma _*$ is the most important parameter in determining the metallicity, followed by the galactocentric distance and $\Sigma _{SFR}$. There is little or no dependence on $\Sigma_{H_2}$ (and, if any, in the opposite direction relative to $\Sigma _{SFR}$), indicating that the inverse correlation of local metallicity with $\Sigma _{SFR}$ is {\it not} a by product of an inverse correlation with $\Sigma_{H_2}$ via the S-K relation.}
    \label{fig:pcc}
\end{figure}

Figure \ref{fig:pcc} further extends this analysis by showing the partial correlation coefficients of the local metallicity with the following quantities: molecular gas surface density ($\Sigma_{H_2}$), stellar mass surface density ($\Sigma_*$), SFR surface density ($\Sigma_{SFR}$), galactocentric distance (R, measured in units of the effective radius), and a uniform random variable (Random).
Once again, we focus for now on the case of the metallicity-dependent $\alpha_{CO}$ conversion factor, which is indicated by the dark blue bars, while we will discuss later the (erroneous) case of a constant $\alpha_{CO}$ in the Appendix.
The PCC analysis clearly shows, once again, that $\Sigma_*$ is the most important parameter in driving the local metallicity. In absence of $M_*$ (which we have shown in Sec. \ref{sec: res and glob}  to be as important as $\Sigma_*$) the predictive power of the latter is taken by the galactocentric distance, which becomes the second most important parameter, possibly as a consequence of the mass-size relation. The most interesting aspect is that the inverse correlation with $\Sigma _{SFR}$ is {\it not} replaced by the $\Sigma _{H_2}$. The dependence on $\Sigma _{H_2}$ is very weak and, if anything, goes in the opposite direction with  respect to the $\Sigma _{SFR}$.

\begin{figure}
    \centering
    \includegraphics[width=\columnwidth]{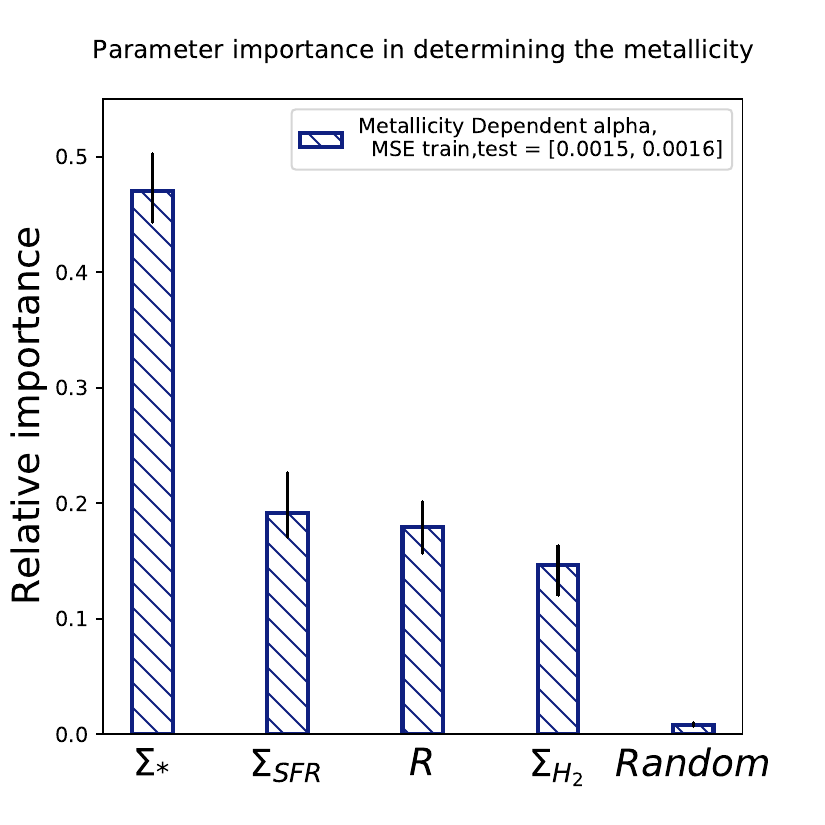}
    \caption{Random Forest regression parameter importance for determining the resolved metallicity in the ALMaQUEST sample. Parameters included are:  stellar mass surface density ($\Sigma_*$),  star formation rate surface density ($\Sigma_{SFR}$),  molecular gas mass surface density ($\Sigma_{H_2}$), galactocentric radius (R), and a random uniform variable (Random). The RF confirms the results of the PCC analysis, i.e. the local metallicity depends primarily on $\Sigma _*$ with a secondary dependence on galactocentric radius and $\Sigma _{SFR}$, whilst a weaker dependence is found with $\Sigma _{H_2}$. }
    \label{fig:RF_alma}
\end{figure}

 Similar results are obtained from the RF analysis, as shown in Figure \ref{fig:RF_alma}, where the bars give the relative importance of the various parameters in driving the local metallicity. Once again, $\Sigma_*$ dominates. Another important parameter is the galactocentric distance, possibly incorporating the predictive power of the total stellar mass (which is absent in this analysis), via the mass-size relationship. The importance of SFR surface density is maintained and {\it not} replaced by $\Sigma _{H_2}$.

 It should be noted that the introduction of the metallicity-dependent $\alpha _{CO}$ might have been a potential worry for introducing artificial correlations, should we have found that $\Sigma _{H_2}$ is strongly correlated with metallicity; however, we find just the opposite, i.e. despite having a metallicity-dependent conversion factor, $\Sigma _{H_2}$ has a very small role in driving the galaxy metallicity. Once again, the case of a constant $\alpha _{CO}$ is discussed in the appendix.

We also note that, as seen earlier, it does appear that there are differences between the random forest regression parameter importances (Figure \ref{fig:RF_alma}) and the strength of the partial correlation coefficients (Figure \ref{fig:pcc}). These relate to the differences between the two methods. The random forest regression can uncover non-monotonic relationships and probes all parameters simultaneously whilst the partial correlation coefficients are limited to monotonic trends. However the partial correlation coefficients enable us to see the sign of the relationship (i.e. whether it is positive or negative) which is not provided by the random forest. This is why we use both methods in our analysis, each method gives us a different piece of information. Our baseline for the strength of the importances, i.e. actual dependencies, is the random forest regression \citep[as it has been shown to be able to uncover intrinsic dependencies amongst highly correlated quantities in ][]{Bluck2022A&A...659A.160B}. We use the partial correlation coefficients to inform us of the direction of the relationship (they also provide a convenient way to check that we uncover similar trends to the random forest). Anyway, it is reassuring that {\it both} methods indicate that, within this set of parameters, $\Sigma_*$ outranks all other quantities, by a significant factor, in predicting the local metallicity.

 Another aspect to consider is whether any of our results may be being influenced by the metallicity tracers used. Our approach mitigates against this issue by using a combination of nine different tracers that probe different metallicity regimes. However,
\cite{Schaefer2020ApJ...890L...3S} and \cite{Schaefer_N/O_2022ApJ...930..160S} have previously highlighted the possible bias in using purely Nitrogen based metallicity indicators. To test against this effect we rerun our main analysis with just the 7 Nitrogen--free metallicity diagnostics. We find that this does not significantly affect our results and any deviations are within the respective errors.

 In Appendix \ref{sec:measurementuncertainty} we test the random forest against the effect of varying measurements uncertainty. This is to ensure it is not being biased by possible greater measurement uncertainty on $\Sigma_{H_2}$ compared to $\Sigma_{SFR}$. We find that even after 5$\sigma$ of Gaussian random noise is added to $\Sigma_{SFR}$, $\Sigma_{H_2}$ is still not responsible for driving the metallicity and the importances of $\Sigma_{H_2}$ and the importance of $\Sigma_{SFR}$ remains similar. This reiterates that the importance of $\Sigma_{SFR}$ in the random forest is not simply tracing $\Sigma_{H_2}$ (and removes the possibility that the importance of $\Sigma_{SFR}$ stems from the possibly reduced measurement uncertainty compared to $\Sigma_{H_2}$).

Our findings indicate that $\Sigma _{SFR}$ is not simply a proxy of $\Sigma_{H_2}$ in driving the metallicity in galaxies, but has a direct role.
In particular, the ``accretion--and--dilution'' scenario for explaining the FMR would predict that the local metallicity is primarily driven by $\Sigma _{H_2}$ and only indirectly  by $\Sigma _{SFR}$. Our finding suggests that the dilution scenario is unlikely to be the only or the primary effect responsible for the inverse, secondary dependence of the metallicity on SFR.

\section{Discussion}

 The two most important parameters for determining the local, resolved metallicity, are the local, resolved stellar mass surface density and the global stellar mass of the entire galaxy.
 The dependence of the local metallicity on local and  total stellar mass has been found previously in \citet{Barrera_global_from_local_radial2016MNRAS.463.2513B}, \cite{Hwang2019ApJ...872..144H}, \citet{Gao2018ApJ...868...89G} and \cite{Boardman22}. However, these previous works did not investigate {\it simultaneously} the dependence of metallicity on $\Sigma _*$, $M_*$ and other galactic quantities, so it was not possible in those earlier works to determine if the dependence on $\Sigma _*$ or $M_*$ was actually a by-product of other correlations. Our analysis has unambiguously determined that the local metallicity depends directly, primarily and independently on both $M_*$ {\it and} $\Sigma _*$.
 
 The dependence of the resolved metallicity on {\it both} $\Sigma _*$ {\it and} $M_*$
can be interpreted as the local metallicity not just resulting from local metal production and retention via the local gravitational potential of the disc, but also global metal production and retention. Redistribution of metals due to radial migration and gas flows (e.g. through stellar bars) is likely responsible for the local metallicity to be associated with the metal production outside the local environment, i.e. across the entire galaxy. Similarly, the global gravitational potential of the entire galaxy must help to retain metals in local regions. \citet{Zibetti2022MNRAS.512.1415Z} find that simultaneously both global and local stellar mass are thought to drive the properties of the stellar population.
 
 Once the role of global and resolved stellar mass is accounted for, there is evidence for a metallicity (inverse) dependence both on local SFR surface density and global SFR of the entire galaxy. These results
 indicate that the existence of the global MZR and FMR cannot be accounted for only as a consequence of the local MZR and FMR, rather that there is a crucial dependence of resolved metallicity also on global stellar mass and global SFR.

The finding that 
the global star formation rate is more important than resolved star formation rate in determining the local metallicity reveals that the ``accretion, dilution and SFR boost'' scenario cannot fully explain the existence of the resolved FMR \citep[as has also been suggested in][]{Hayden_pawson2022MNRAS.512.2867H}. In the latter scenario the accretion of metal-poor (near-pristine) gas would result in locally decreasing the metallicity and locally increasing the SFR as a result of the freshly supplied fuel for star formation: in this case we would expect that the local metallicity should only anti-correlate with the local $\Sigma _{SFR}$, in contrast to the finding that the dependence is stronger with the total, global SFR.
 It remains possible that the gas could come from a small dwarf galaxy falling into the larger galaxy, being re-distributed kinematically across the host galaxy, which could then enhance the spiral structure and result in a global SFR boost. This could account for the dependence on global SFR.

The fact that ``accretion, dilution and SFR boost''  cannot fully explain the rFMR is further supported by the results obtained with the ALMaQUEST data. In the ``accretion, dilution and boost'' scenario we would expect $\Sigma _{SFR}$ to be just a proxy of $\Sigma _{H_2}$ via the S-K relation, and that the primary, intrinsic dependence of the metallicity should be on the local surface density of molecular gas. The fact that when $\Sigma _{H_2}$ is introduced in the sample this does {\it not} become the primary driving parameter of the metallicity, i.e. it does {\it not} replace $\Sigma _{SFR}$, provides further argument against the simple ``accretion, dilution and SFR boost'' scenario. The accretion dilution-and-star formation boosting scenario likely contributes to the rFMR, but cannot be the only explanation.

Therefore, both SFR and $\Sigma _{SFR}$ must have a direct role in locally reducing the metallicity. 
 The inverse dependence on global star formation could be linked to ejection of metals by galactic-scale (metal-loaded) winds caused by the combination of SNe across most of the galaxy; such galactic-scale winds would affect the metallicity of regions across most of the galaxy.  
The dependence on local star formation may be via local SF-driven, metal-loaded winds, which preferentially eject metals from the disc \citep[][]{Konami_metal_loaded_outflow2011PASJ...63S.913K, Origlia_metal_loaded_winds2004ApJ...606..862O}.
 Indeed, SN-driven galactic winds are known to be more metal rich than the ISM of the galaxy they develop from, as they carry significant amounts of the freshly enriched gas produced by the SNe explosions \citep{Origlia_metal_loaded_winds2004ApJ...606..862O, Konami_metal_loaded_outflow2011PASJ...63S.913K, Chisholm2018MNRAS.481.1690C}.
  Metal rich winds have also been found from a theoretical perspective to be required to explain the scaling relations of metallicity gradients \citep[][]{Sharda_gradients_2021MNRAS.502.5935S, Sharda_metallicity_gradient_winds2021MNRAS.504...53S}.

However, there would appear to be issues with winds as an explanation for the local SFR importance. Indeed, we note that in a starburst event many of the newly synthesised metals will be in a hot phase (of the order of $10^8$ K) which would be ejected from the disc more easily \citep[][]{Yates2020A&A...634A.107Y}. This hot gates cools and mixes into the galaxy on very long timescales ($\sim$100~Myr--Gyr) in the halo, hence the resulting metal deficiency is unlikely to be traced until after the H$\alpha$ SFR timescale. However the hot gas cooling times are much shorter in the disc, due to shocking with denser gas. Due to these timescale aspects, metal loaded galactic winds may potentially explain the metallicity inverse dependence on {\it global} SFR, which is associated with longer timescales, but are less likely to explain the local anti-correlation with $\Sigma _{SFR}$. 

 Other speculative possibilities include (but are not limited to) stellar M/L ratios being biased for spaxels with high SFR, a metallicity-dependent star formation efficiency \citep[][]{Dib_zdepSFE_2011MNRAS.415.3439D}, or radiation pressure on dusty clouds, which would preferentially ejected metal richer ISM.
We also reiterate that these are just possible interpretations of our results and more detailed investigations are needed with theoretical models and numerical simulations. 

It is important to compare our ALMaQUEST findings with previous studies that have investigated the dependence of metallicity with the gas content of galaxies. As mentioned in the introduction, past studies \citep[][]{Bothwell2016A&A...595A..48B,Brown2018MNRAS.473.1868B, Chen_HI_2022arXiv220508331C} have found clear evidence for an inverse correlation between {\it global} metallicity and {\it global} gas content, with lower dispersion relative to the inverse correlation with SFR, suggesting that the FMR actually is a by-product of a more fundamental (anti-)correlation with the gas content, supporting the dilution scenario as primary origin of the FMR. These past results seem at odds with our reported findings. However, one should take into account that those results are based on the {\it total} gas mass (and global metallicity), which we cannot test with the limited ALMAQUEST sample (although a study extending this work to the integrated gas properties, by using additional samples, is in progress). More importantly, the bulk of those results are based on atomic HI gas mass, and not molecular gas \citep[the studies using molecular gas had much lower statistics][]{Bothwell2016MNRAS.455.1156B}. In contrast to the molecular gas mass, the atomic gas does not participate in the star formation and is distributed on much larger scales. Therefore it is possible that the anti-correlation of the global metallicity (possibly dominated by the external regions) with the global HI is indeed strong and supporting a dilution scenario for the galactic discs on large scales, while in the inner, more active regions, the contribution of metal expulsion by SF plays an additional role. Exploring these differences further will require both large samples of spatially resolved HI maps, and large samples of galaxies with integrated molecular gas masses and resolved metallicity information.

Finally, the finding that, once the dependence on local and global quantities is taken into account, the local metallicity has a residual dependence on the galactocentric distance (although weak), indicates that metallicity radial gradients are not only an indirect, byproduct of local scaling relations, but must also have an intrinsic nature. More specifically, the scenario in which the metallicity negative radial gradient is the result of the resolved MZR combined with the radially declining $\Sigma _*$ may be a plausible, primary explanation, but cannot fully account for the metallicity gradient. The intrinsic, negative dependence found by us of the local metallicity on galactocentric distance, may result from inward radial migration of metals, expected by various models \citep{Spitoni2015MNRAS.451.1090S}, possibly also associated with inflows driven by stellar bars or galactic fountains. Another possibility would be the migration of metal-poor (i.e. almost pristine) HI from the outer disc (beyond star forming regions), travelling in and lowering the metallicity at large galactocentric radii. Such kind of radial motions in discs have been suggested previously \citep[][]{Sanchez_discs_oxygen_2014A&A...563A..49S,Tissera_2019MNRAS.482.2208T,Yates2021MNRAS.503.4474Y}

\section{Conclusions}

Using data from the MaNGA and the ALMaQUEST surveys, we applied partial correlation coefficients and random forest regression to explore metallicity dependencies on both resolved and global properties. This enabled us to investigate several inter-correlated quantities and determine what are their roles in determining the local metallicity and, in particular, cleanly disentangle intrinsic, direct correlations from those dependencies that are indirect.

Our are main observational findings are the following:
\begin{itemize}
    \item We unambiguously confirm the existence of a resolved fundamental metallicity relation (rFMR): the local metallicity primarily depends on the local stellar mass surface density ($\Sigma _*$) and has a secondary (inverse) dependence on the local surface density of star formation rate ($\Sigma _{SFR}$).
    \item However, when also combined with global properties, the resolved metallicity depends on both global and local properties. This indicates that the global MZR and FMR do not simply result from the resolved MZR and FMR.
    \item The primary two metallicity drivers are $\Sigma _*$ and $M_*$, which play a similar role in determining the local metallicity.
    \item A secondary role in driving the local metallicity is played by the star formation rate. We find that the global SFR is a more important parameter than $\Sigma_{SFR}$ in determining the resolved metallicity. We also find that the importance of $\Sigma_{SFR}$ decreases in the highest mass galaxies.
    \item The molecular gas content $\Sigma _{H_2}$ is less important than $\Sigma_{SFR}$ in regulating the metallicity. 
    \item In addition, once all resolved and global dependencies are taken into account, there is a residual, intrinsic dependence of the local metallicity
    on galactocentric distance, implying that the radial metallicity gradients are not simply a consequence of the spatially resolved metallicity scaling relations (in particular of the $\Sigma_*$-Z relation) but also have an intrinsic component.
\end{itemize}

From these observational results we infer the following implications for our understanding of the mechanisms driving the scaling relations:

\begin{itemize}

\item The metallicity dependence on both local $\Sigma _*$ and M$_*$ implies that metallicity is not driven only by the local metal production and gravitational potential, but also by the global metal production (e.g. via mixing, migration and feedback effects) and the global gravitational potential well.

\item The fact that the local metallicity (inversely) depends more strongly on the global SFR then the local $\Sigma _{SFR}$, as well as the lack of significant dependence on $\Sigma _{H_2}$,
indicates that ``accretion, dilution and SFR boost'' scenario alone cannot explain the existence of the FMR.  The importance of SFR is possibly also linked to the ejection of metals by star forming galactic winds, which are caused by the cumulative result of SNe.

\item Radial migration and transport of metals must play a direct role in shaping the metallicity gradients, in addition, the latter being a by-product of the spatially resolved scaling relations.

\end{itemize}

Finally for the primary scaling relations that we have identified, and particularly the rFMR and the metallicity--$\Sigma _*$-$M_*$ relation, we provide parametrisations of the rFMR in terms of fitting with a 3D surface and projections that minimise the scatter, which can be useful for comparison with other samples, especially at high redshift, for exploring any evolution of these properties.

\section*{Acknowledgements}

W.B., R.M., M.C, and A.B. acknowledge support by the Science and Technology Facilities Council (STFC) and ERC Advanced Grant 695671 "QUENCH". RM also acknowledges funding from a research professorship from the Royal Society.

HAP acknowledges support by the Ministry of Science and Technology of Taiwan under grant 110-2112-M-032-020-MY3.

The authors would like to thank the staffs of the East-Asia and North-America ALMA ARCs for their support and continuous efforts in helping produce high-quality data products. This paper makes use of the following ALMA data:\\
ADS/JAO.ALMA\#2015.1.01225.S,
ADS/JAO.ALMA\#2017.1.01093.S, 
ADS/JAO.ALMA\#2018.1.00541.S,
\\and ADS/JAO.ALMA\#2018.1.00558.S. 

ALMA is a partnership of ESO (representing its member states), NSF (USA) and NINS (Japan), together with NRC (Canada), MOST and ASIAA (Taiwan), and KASI (Republic of Korea), in cooperation with the Republic of Chile. The Joint ALMA
Observatory is operated by ESO, AUI/NRAO and NAOJ.

Funding for the Sloan Digital Sky Survey IV has been provided by the Alfred P. Sloan Foundation, the U.S. Department of Energy Office of Science, and the Participating Institutions. SDSS acknowledges support and resources from the Center for High-Performance Computing at the University of Utah. The SDSS web site is www.sdss.org.

SDSS is managed by the Astrophysical Research Consortium for the Participating Institutions of the SDSS Collaboration including the Brazilian Participation Group, the Carnegie Institution for Science, Carnegie Mellon University, Center for Astrophysics | Harvard \& Smithsonian (CfA), the Chilean Participation Group, the French Participation Group, Instituto de Astrofísica de Canarias, The Johns Hopkins University, Kavli Institute for the Physics and Mathematics of the Universe (IPMU) / University of Tokyo, the Korean Participation Group, Lawrence Berkeley National Laboratory, Leibniz Institut für Astrophysik Potsdam (AIP), Max-Planck-Institut für Astronomie (MPIA Heidelberg), Max-Planck-Institut für Astrophysik (MPA Garching), Max-Planck-Institut für Extraterrestrische Physik (MPE), National Astronomical Observatories of China, New Mexico State University, New York University, University of Notre Dame, Observatório Nacional / MCTI, The Ohio State University, Pennsylvania State University, Shanghai Astronomical Observatory, United Kingdom Participation Group, Universidad Nacional Autónoma de México, University of Arizona, University of Colorado Boulder, University of Oxford, University of Portsmouth, University of Utah, University of Virginia, University of Washington, University of Wisconsin, Vanderbilt University, and Yale University.

\section*{Data Availability}

The ALMA data used is publicly available through the ALMA archive http://almascience.nrao.edu/aq/.

The MaNGA data that is used in this work is publicly available at 
https://www.sdss.org/dr15/manga/manga-data/.

The MPA-JHU catalogue is publicly available at https://wwwmpa.mpa-garching.mpg.de/SDSS/DR7/.

The NASA Sloan Atlas is available at http://nsatlas.org/data.



\bibliographystyle{mnras}
\bibliography{example} 




\appendix

\section{Effect of aperture size differences between SDSS DR7 and MaNGA}
\label{sec:app_test}
An important effect to test against is whether aperture differences between the SDSS DR7 data and the MaNGA data could be biasing the results of section \ref{sec:SDSSvMaNGA}.  For the SDSS data the aperture size is the 3'' probed by the fibre. This will correspond to a different physical aperture size for each galaxy depending on its distance.
This means that for the most nearby galaxies in SDSS the fibre is only sampling the central region of the galaxy. 

\begin{figure}
    \centering
    \includegraphics[width=\columnwidth]{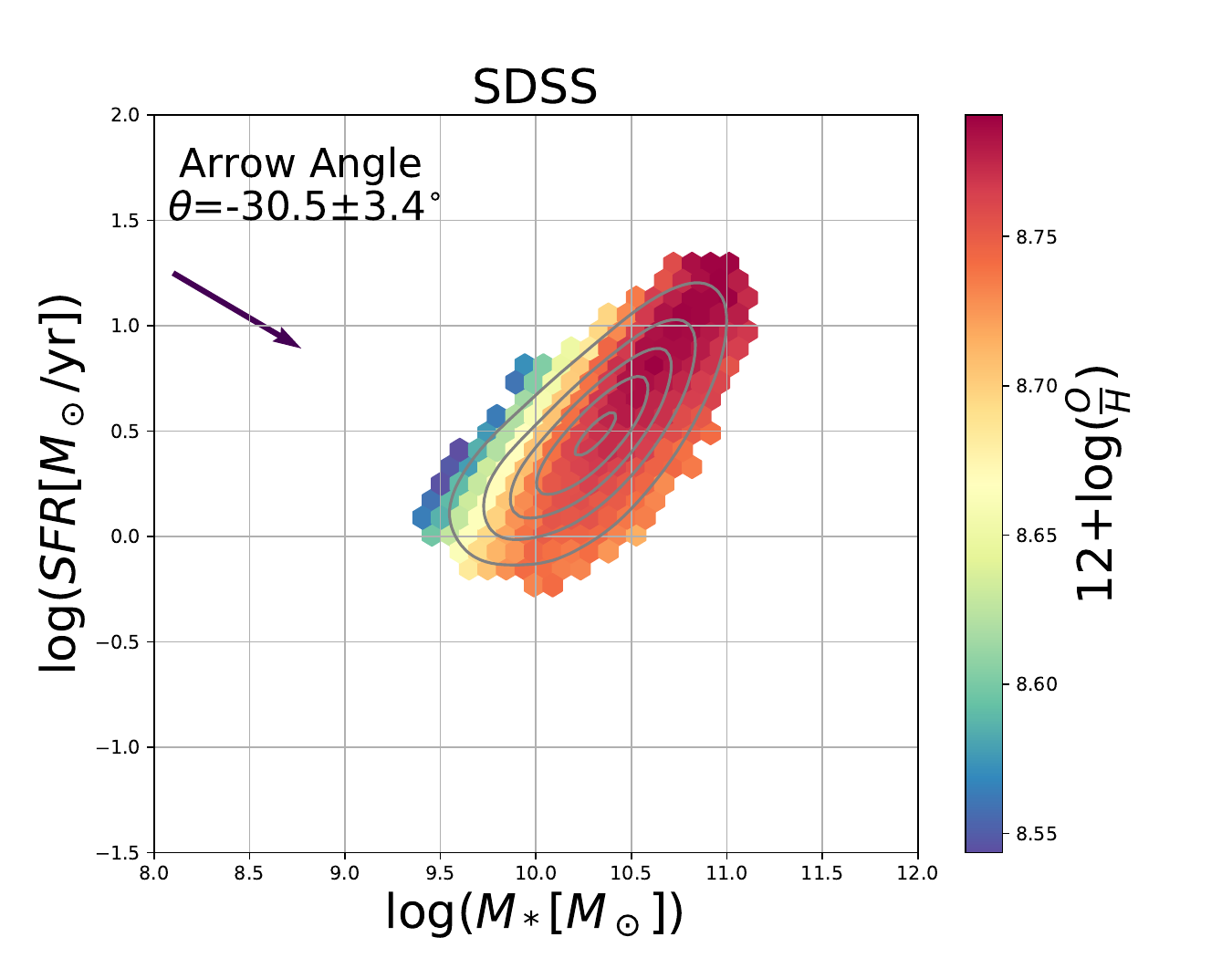}
    \caption{2D histogram of star formation rate against stellar mass, colour coded by the mean metallicity for galaxies in SDSS DR7. The plot is similar to the lower SDSS plot in  Figure \ref{fig:global_fmr}, for a redshift limited range of 0.07<z<0.30. As can be seen, this primarily removes local low stellar mass and low SFR galaxies. It can be seen that the arrow angle remains similar to before, meaning that the fundamental metallicity relation remains even after excluding the most local galaxies.}
    \label{fig:SDSS_redshift_cut}
\end{figure}

Therefore, we test the effect of excluding galaxies with redshifts outside the range 0.07<z<0.30 from the SDSS sample \citep[as has been used previously in][]{2010Mannucci}. 
We plot the result of this redshift cut in Figure \ref{fig:SDSS_redshift_cut}, where the plot is exactly the same as in Figure \ref{fig:global_fmr} lower, except with a reduced number of galaxies.  We note that the redshift cut has primarily removed lower mass galaxies since these objects only meet the minimum flux threshold for being targeted by SDSS at lower redshifts. 
We can see from this Figure that we recover the FMR and its shape is similar to before. This suggests that Figure \ref{fig:global_fmr} is not being significantly biased by the most local galaxies.

Comparing the aperture size between the single SDSS fibre and the effective aperture used to calculate the metallicity in MaNGA is also important.  Sloan metallicities are calculated using the same diagnostics as in MaNGA but the aperture size is 3''. Incorporating the previous redshift cut, this then gives us a mean physical aperture size of 2.9kpc and a median of 3.1kpc. In MaNGA we take the average metallicity of the spaxels within the effective radius of the galaxy. We find that this gives a mean effective physical aperture size of 3.4kpc and a median of 3kpc. This means that the average physical aperture size is similar between the two surveys hence aperture effects do not play a significant role in these results.

\section{Selection differences between ALMaQUEST and MaNGA}
\label{sec:alvman}
\begin{figure}
    \centering
    \includegraphics[width=\columnwidth]{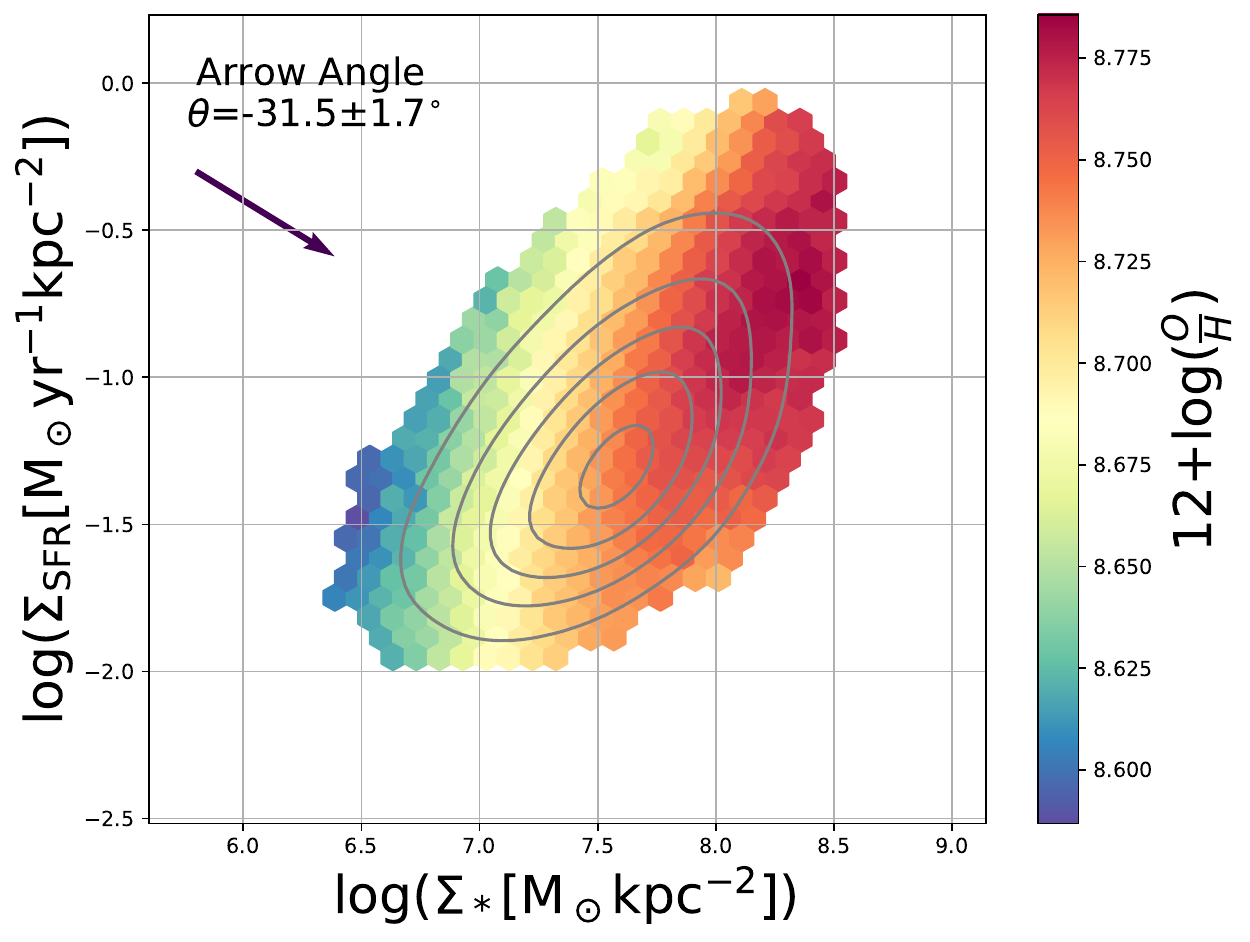}
    \caption{Same as Figure \ref{fig:resolved_FMR_manga} (i.e. shows resolved FMR in MaNGA) but only including spaxels belonging to galaxies within the stellar mass range $10<\rm log(M_*/M_\odot)<11.6$ in order to match the mass ranges of galaxies in the ALMaQUEST survey. Can see that the arrow angle in this scenario closely matches that of the ALMaQUEST rFMR plot, left-hand-side Figure \ref{fig:M-S-O-Alma}. This means that the rFMR seen in the ALMaQUEST sample is in good agreement with the rFMR found in the MaNGA sample.}
    \label{fig:HeavymassFMRManga}
\end{figure}

In Figure \ref{fig:resolved_FMR_manga} we revealed the resolved fundamental metallicity relation in the MaNGA sample and in Figure \ref{fig:M-S-O-Alma} the subset that is the ALMaQUEST sample. We note that the difference in the arrow angle between the two is ~9$^\circ$. We believe this is likely to be caused by the ALMaQUEST survey sampling distribution being skewed towards higher stellar masses than the MaNGA sample. The final ALMaQUEST spaxels (after all cuts etc) belong to galaxies ranging in log stellar mass from 10 to 11.6. To test whether this is the case we use a reduced subset of the MaNGA sample containing spaxels that belong to galaxies of stellar masses within the same range as those in ALMaQUEST. This removes spaxels of low mass galaxies. 
In Figure \ref{fig:HeavymassFMRManga} we re-plot Figure \ref{fig:resolved_FMR_manga}, for this reduced subset of the MaNGA sample. As can be seen with this stellar mass range in place, the arrow angle in the MaNGA sample becomes $\theta=-31.5^\circ$ compared to the $\theta=-31.3^\circ$ in ALMaQUEST. 
This means that the ALMaQUEST sample is representative of the MaNGA sample, in particular for galaxies of higher stellar mass. 

\section{Further surface plots}
\begin{figure}
    \centering
    \includegraphics[width=\columnwidth]{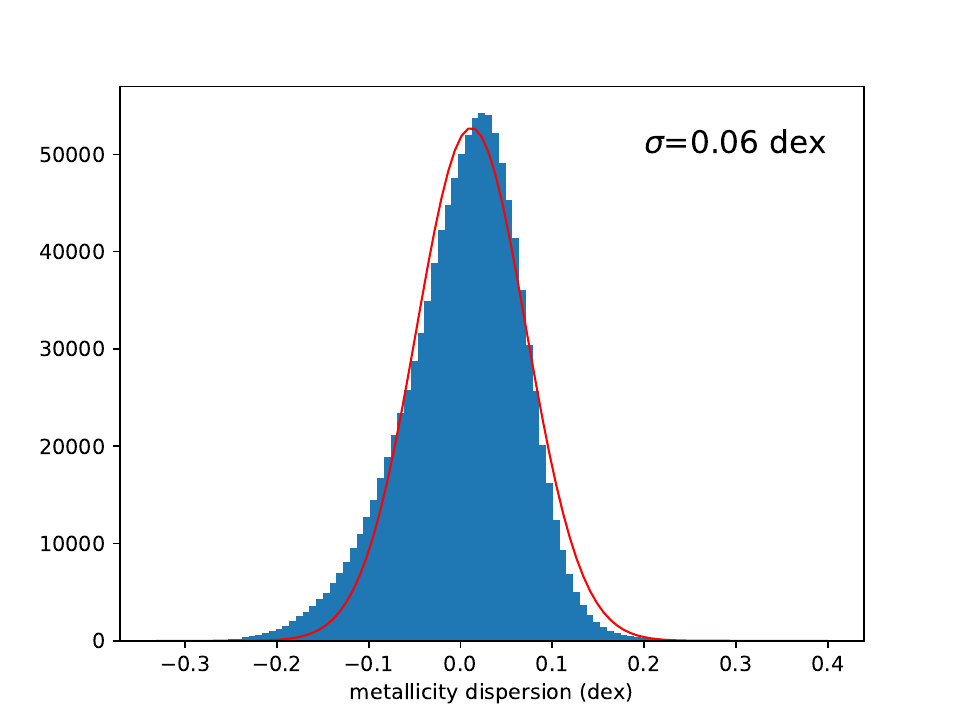}
    \caption{Individual spaxels' metallicity dispersion around the rFMR. The red line is a gaussian fit to the histogram with a standard deviation of $\sigma=0.60$dex. This shows that individual spaxels follow a tight relation around the rFMR as parameterised in Table \ref{table:fits}.}
    \label{fig:dispersion}
\end{figure}

In section \ref{sec:param} we fit a 2D surface to bins of $\Sigma_*$ and $\Sigma_{\rm SFR}$.
Figure \ref{fig:dispersion} shows the a histogram of the metallicity dispersion of the individual  spaxels with the fitted surface. The red line is a gaussian fit to the histogram with standard deviation $\sigma=0.06$ dex. Hence, we obtain a tight scatter around this new parameterisation of the rFMR, where the dispersion is similar to that found by \citet{2010Mannucci} for the integrated SDSS galaxies. 

\begin{figure}
    \centering
    \includegraphics[width=0.9\columnwidth]{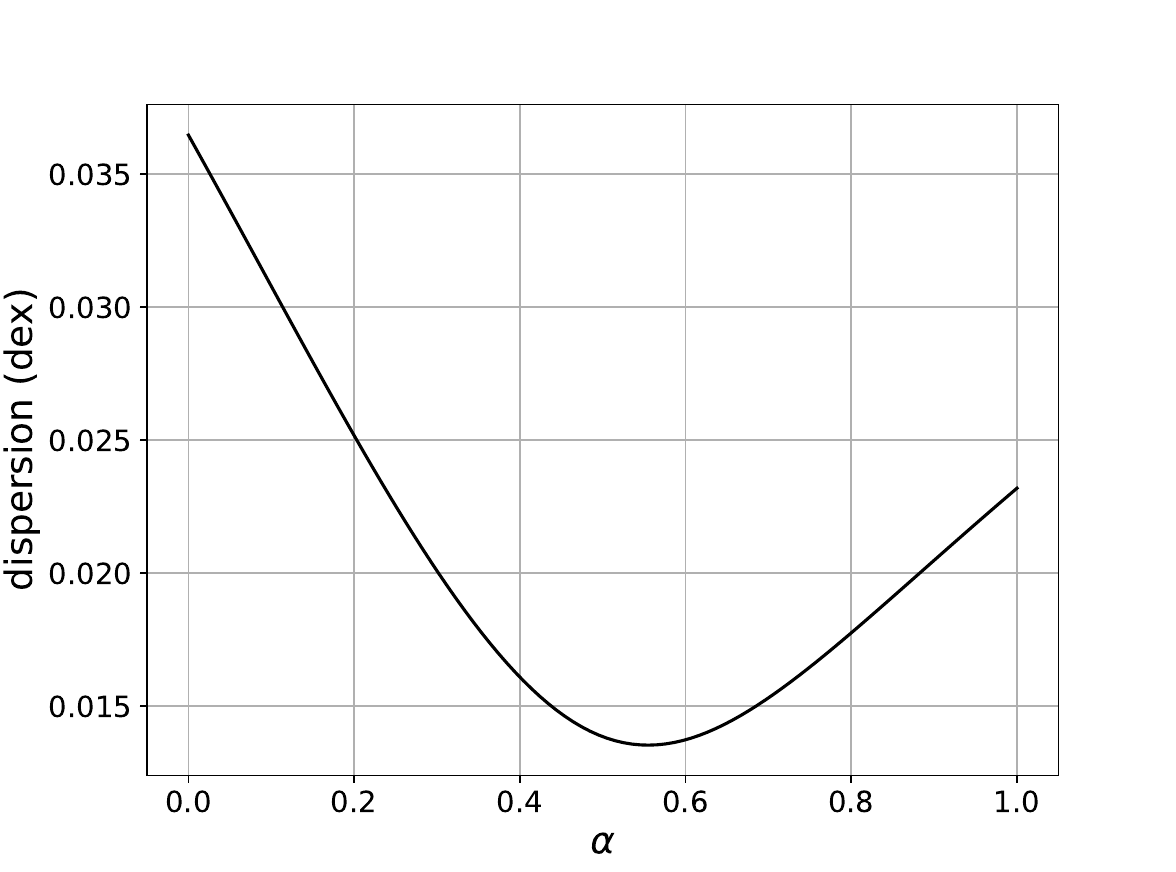}
    \caption{Dispersion of the (binned) metallicity from the best-fit, calculated in bins of $\mu _{\alpha}$, where the latter is defined in Equation \ref{eq:mu_alpha}. The minimum dispersion is found for the value $\alpha=0.54$, meaning that to minimise the scatter of the metallicity to the fit, a combination of both $\Sigma_*$ and $\Sigma_{\rm SFR}$ are necessary.  }
    \label{fig:alpha}
\end{figure}
In section \ref{sec:param} we explored which representation of the 3D plot minimises the scatter.
We include here, Figure \ref{fig:alpha}, which shows $\alpha$ against the mean dispersion of the metallicity. We see that for the resolved data the minimum (turning point) is given by a value of 0.54. This is the value of $\alpha$ for which the scatter is minimised.

\begin{figure*}
    \centering
    \includegraphics[width=1\columnwidth]{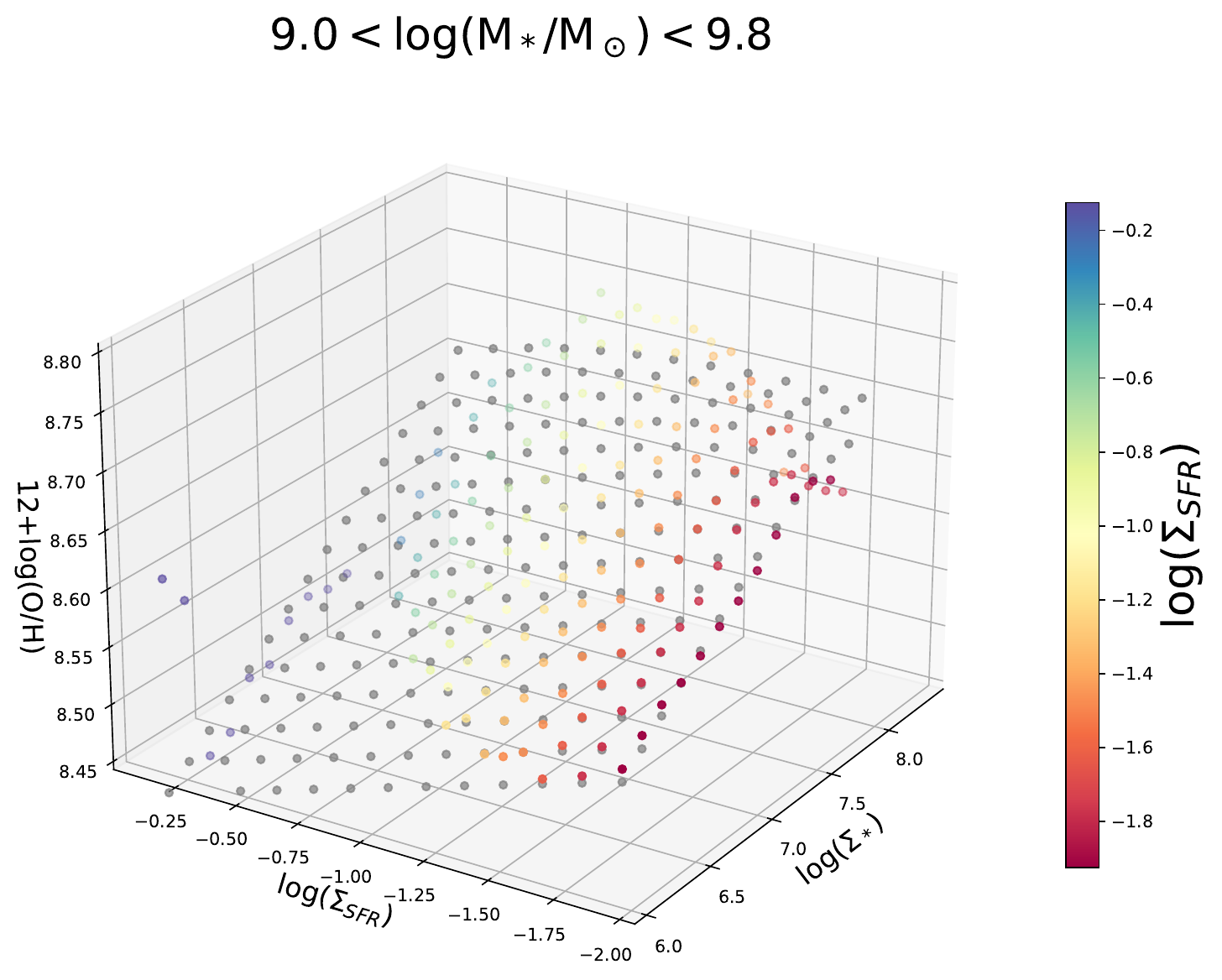}
    \includegraphics[width=1\columnwidth]{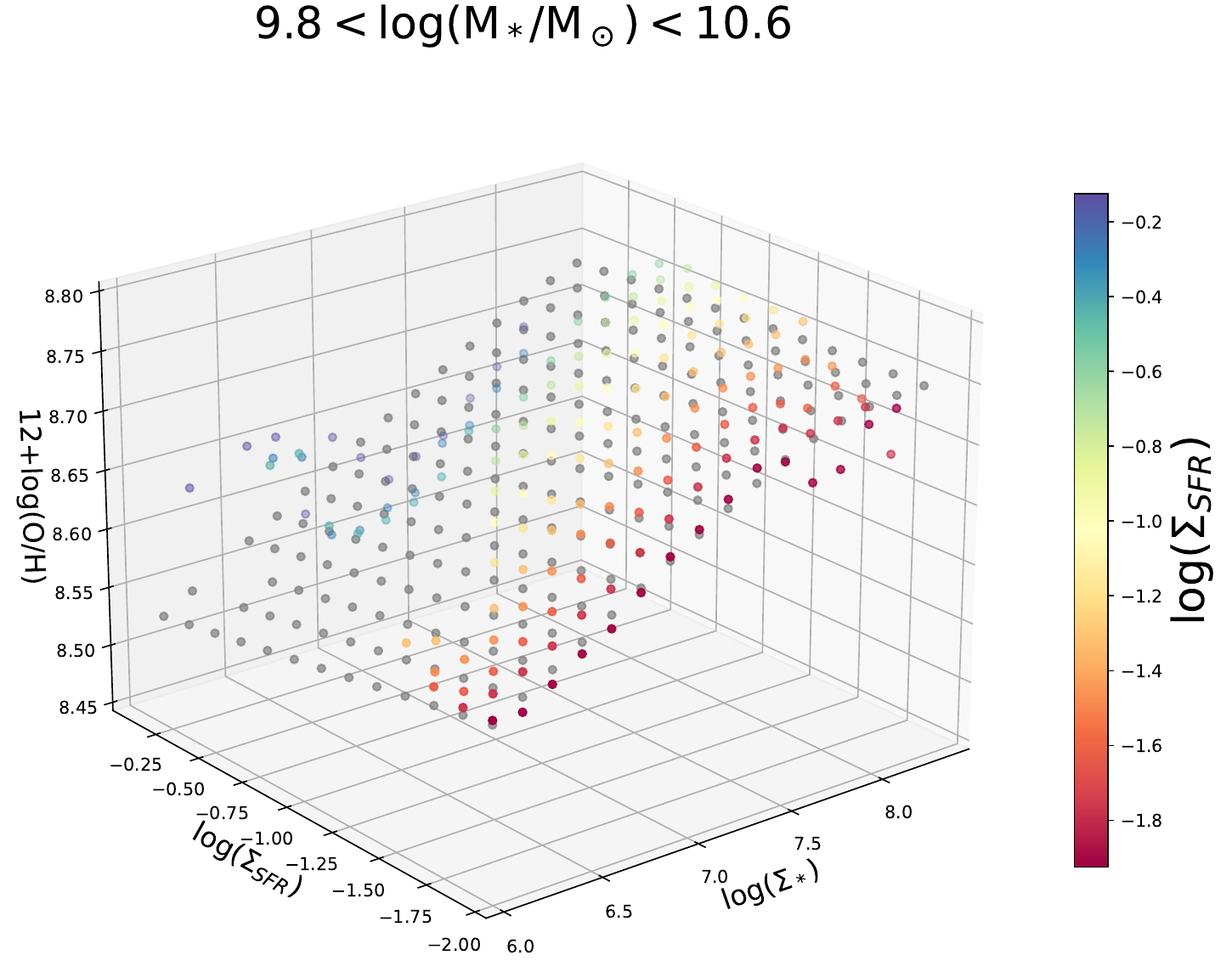}
    \includegraphics[width=1\columnwidth]{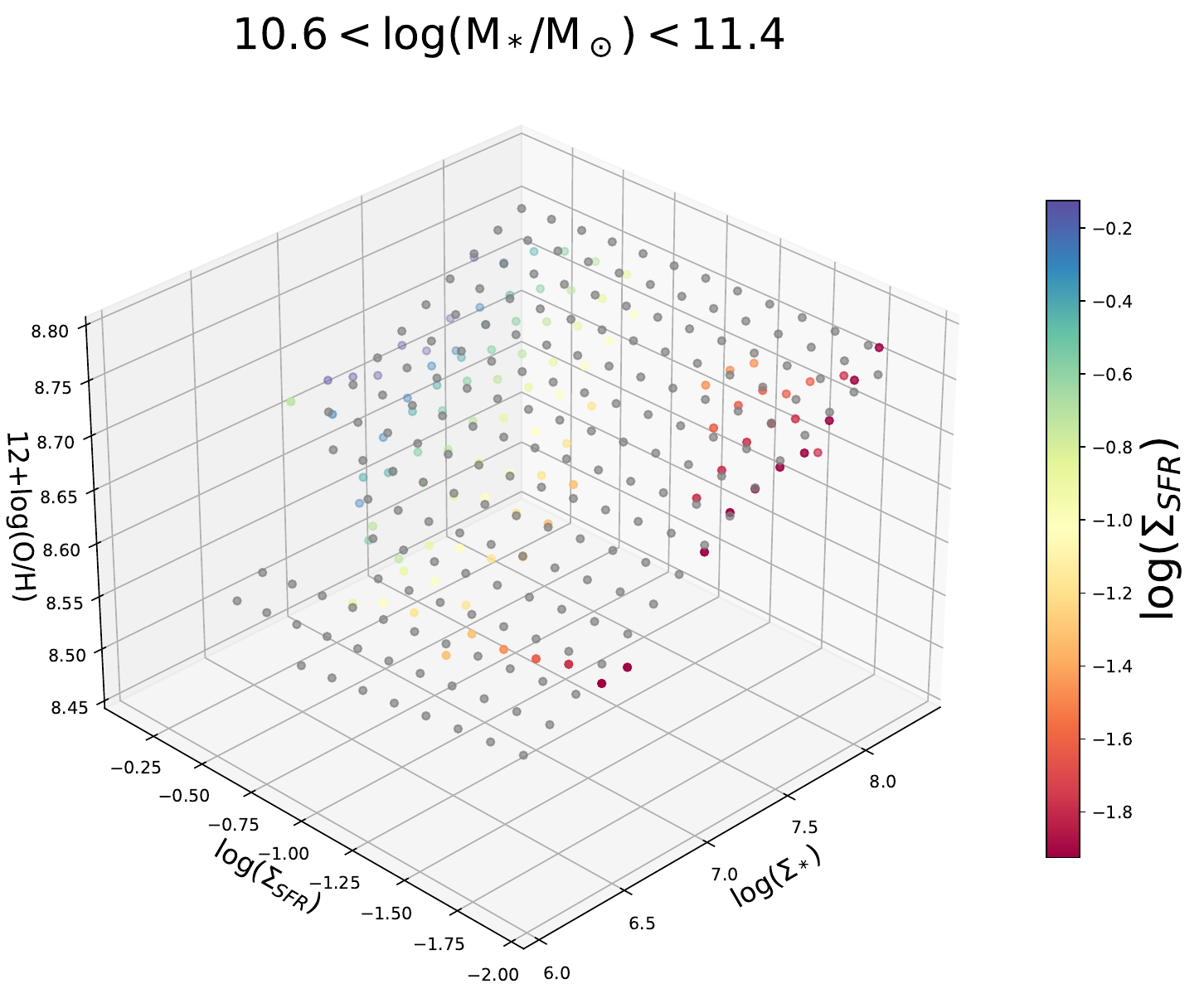}
    
    \caption{3D parametrisation of the resolved FMR in three bins of stellar mass for: top left, low stellar mass bin; top right, medium stellar mass bin; and bottom, the high stellar mass bin. The coloured points are bins of $\Sigma_*-\Sigma_{\rm SFR}$ with the mean metallicity per bin. The grey points are the best-fit surface to the bins. The bins follow the surface well although there are deviations at the edges. The surface is shifted to higher metallicities as the stellar mass bin increases. This offers further proof of the importance of total stellar mass in determining the resolved metallicity.}
    \label{fig:fmr_param_bins}
\end{figure*}

We also explored different parameterisations for the varying bins of total stellar mass (see Table \ref{table:fits}). Figure \ref{fig:fmr_param_bins} shows the resulting 3D plot for: left, 9.0<log($M_*$/$M_\odot$)<9.8; right, 9.8<log($M_*$/$M_\odot$)<10.6; and bottom, 10.6<log($M_*$/$M_\odot$)<11.4. We can see that the surfaces fit well, although can have issues at the edges (as is the case for the main plot).

\begin{figure*}
    \centering
    \includegraphics[width=1\columnwidth]{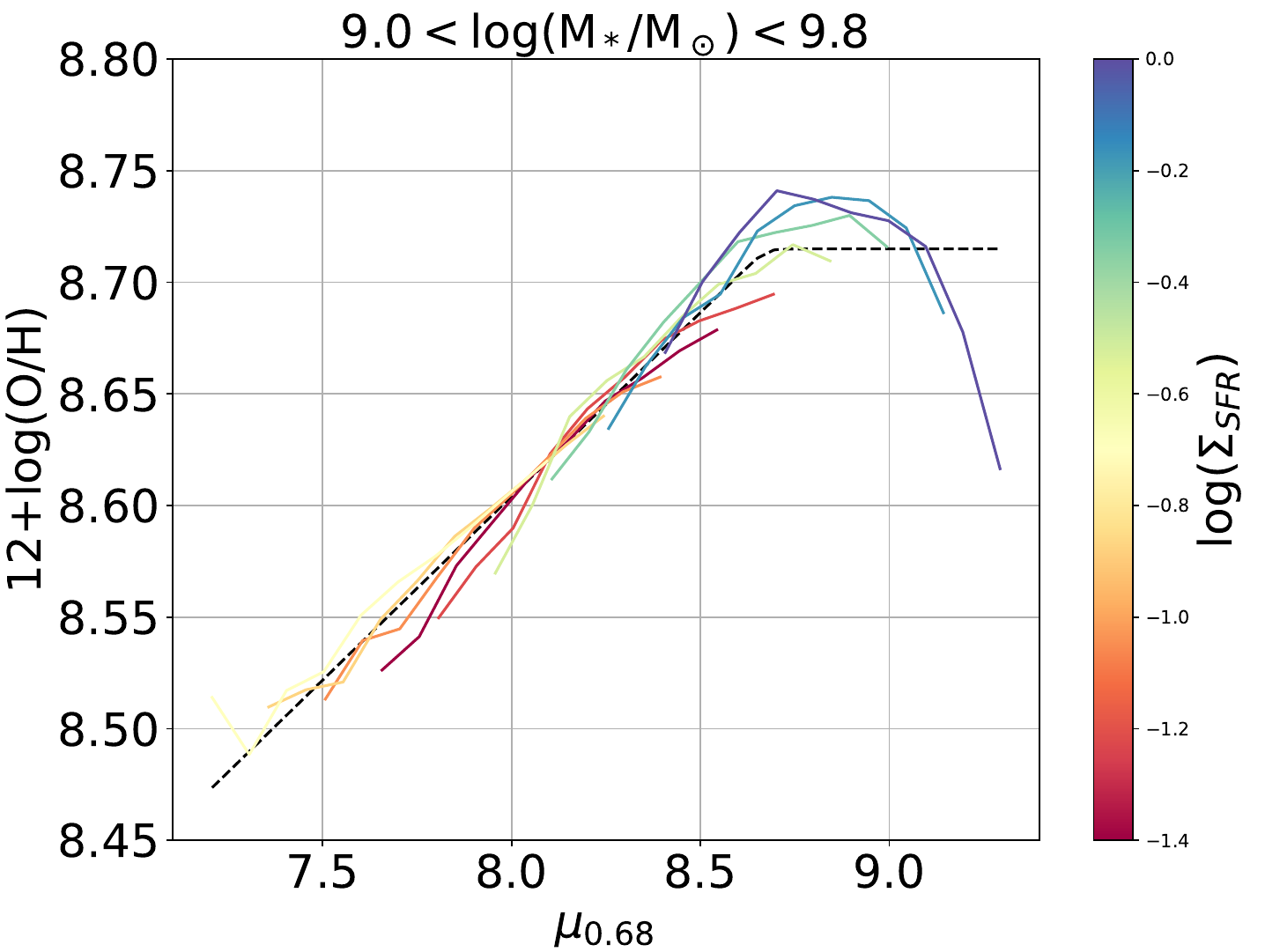}
    \includegraphics[width=1\columnwidth]{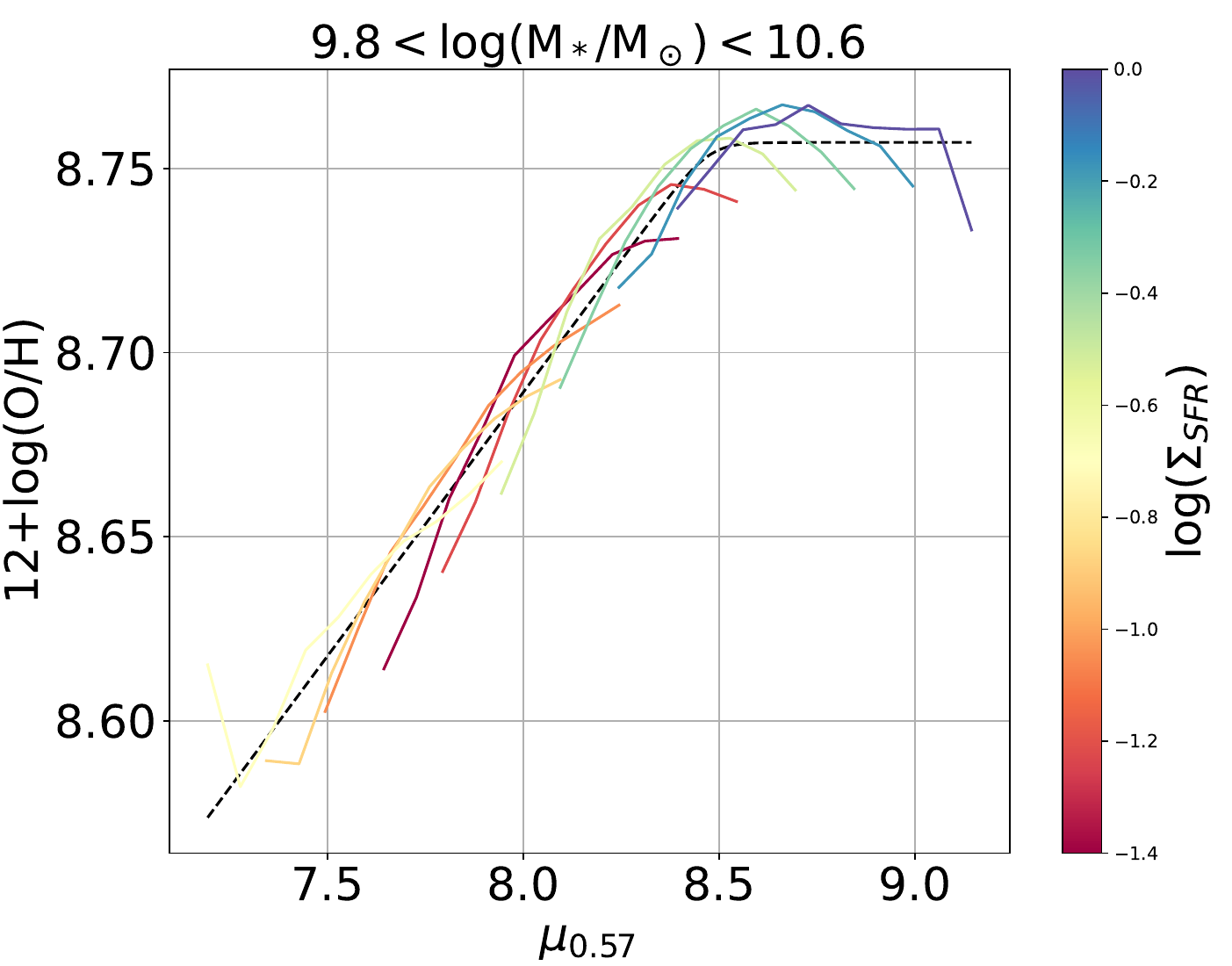}
    \includegraphics[width=1\columnwidth]{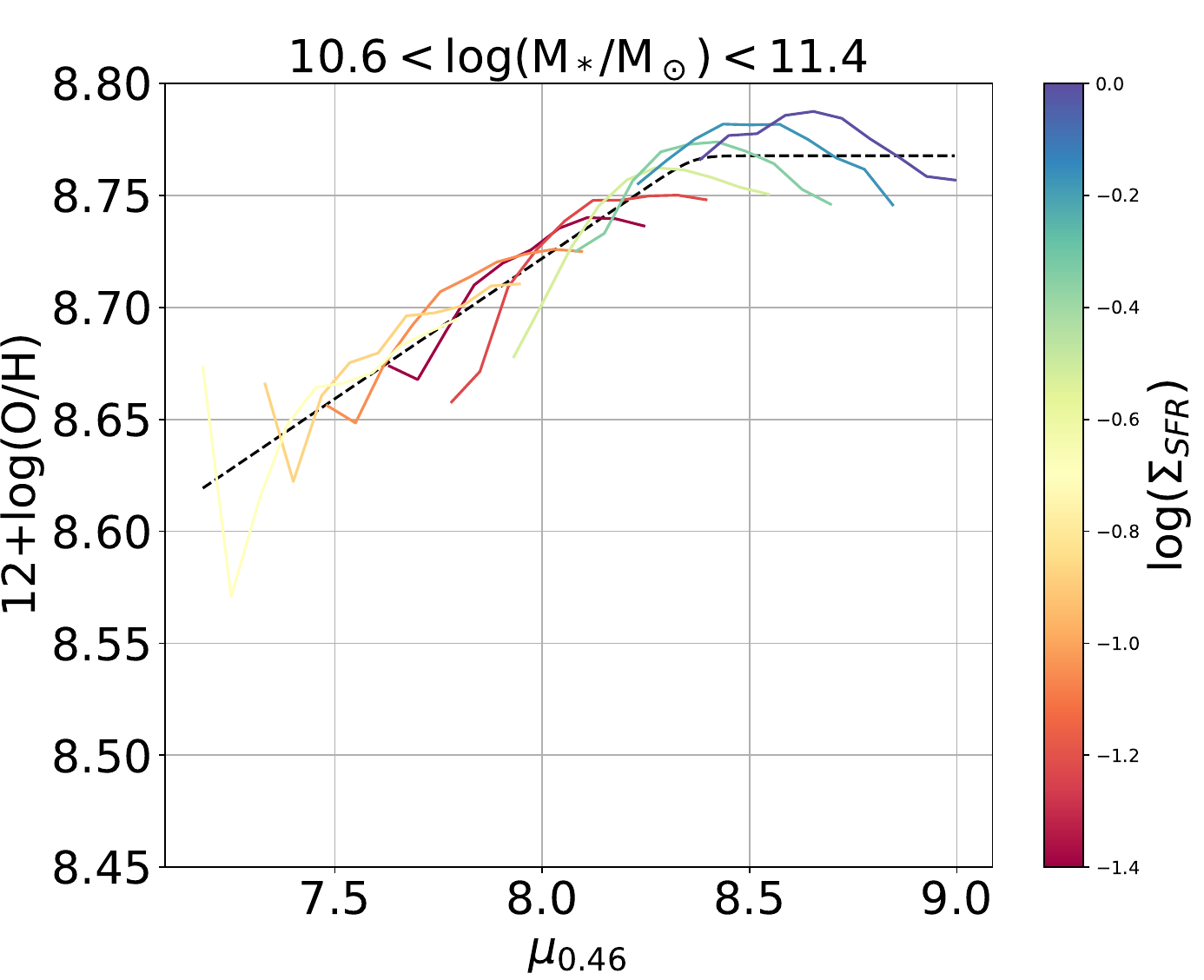}
    
    \caption{Metallicity versus $\rm \mu _{\alpha}= log(\Sigma _*) -\alpha log(\Sigma _{SFR})$ for the values of $\alpha$ that minimise the scatter for: top left, the low total stellar mass bin, top right: the medium total stellar mass bin, and bottom: the high total stellar mass bin. The tracks show minimal dispersion about the best fit. Can see how, at a given fixed $\mu$, the tracks evolve with the increasing stellar mass bins to higher metallicity between the diagrams, i.e. for higher mass bins of $\rm log(M_*)$ the tracks are offset to greater metallicities. }
    \label{fig:best_alpha_bins}
\end{figure*}

In addition, we were able to find the projection of the the 3D plot that reduces the scatter between the bins. We did this by defining $\rm \mu=log(\Sigma_*)-\alpha\,log(\Sigma_{SFR})$ (a combination of resolved stellar mass and star formation rate).  This gave us a value of $\alpha$ which describes the relative contribution of log($\Sigma_{\rm SFR}$) to the best projection. We repeated this for the total stellar mass bins and found the respective values of $\alpha$ (given in table \ref{table:fits}). Here we include, in Figure \ref{fig:best_alpha_bins}, the three respective plots each showing binned metallicity versus $\mu_{\alpha}$ in tracks of $\Sigma_{\rm SFR}$. It can be seen that the tracks in each plot tightly follow the best fit lines with minimal separation.

\section{Parametrisation of the dependence on the total stellar mass}

\begin{figure*}
    \centering
    \includegraphics[width=2\columnwidth]{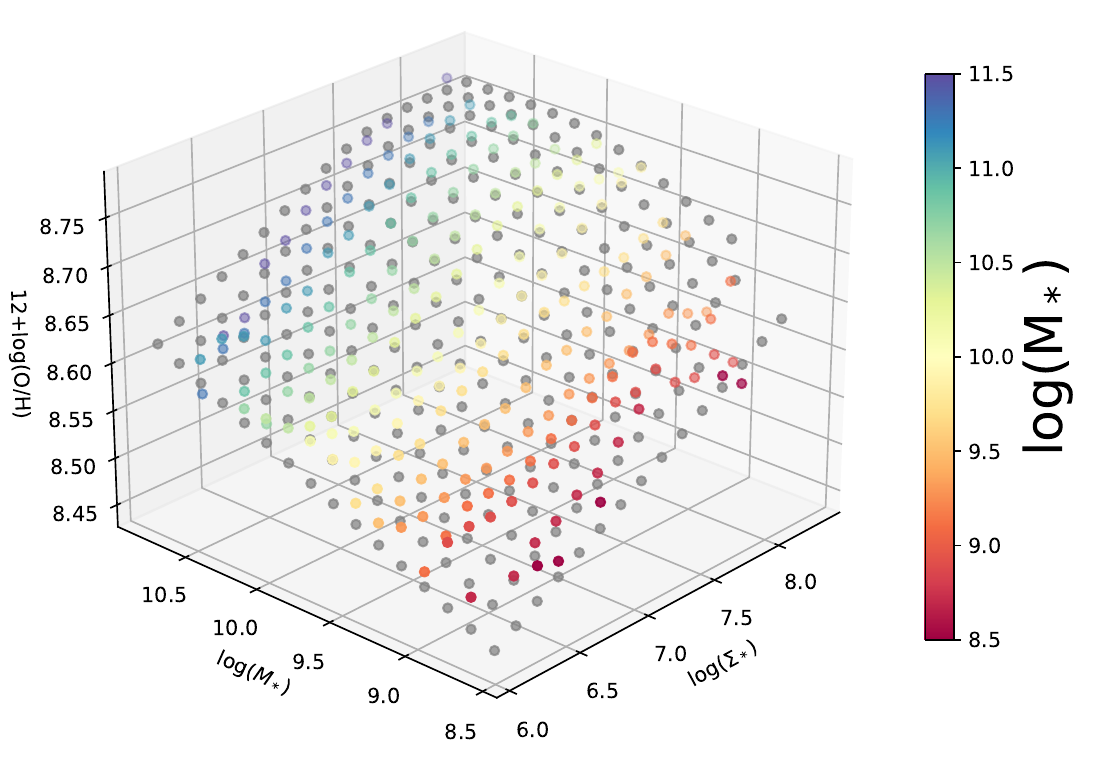}
    \includegraphics[width=1.1\columnwidth]{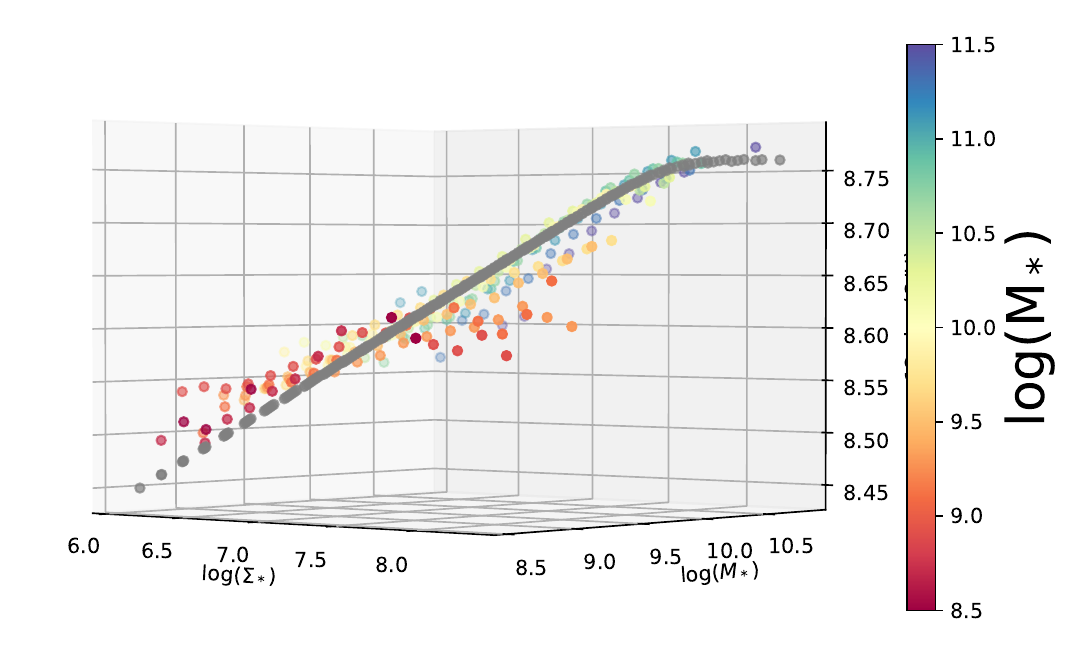}
    \includegraphics[width=0.9\columnwidth]{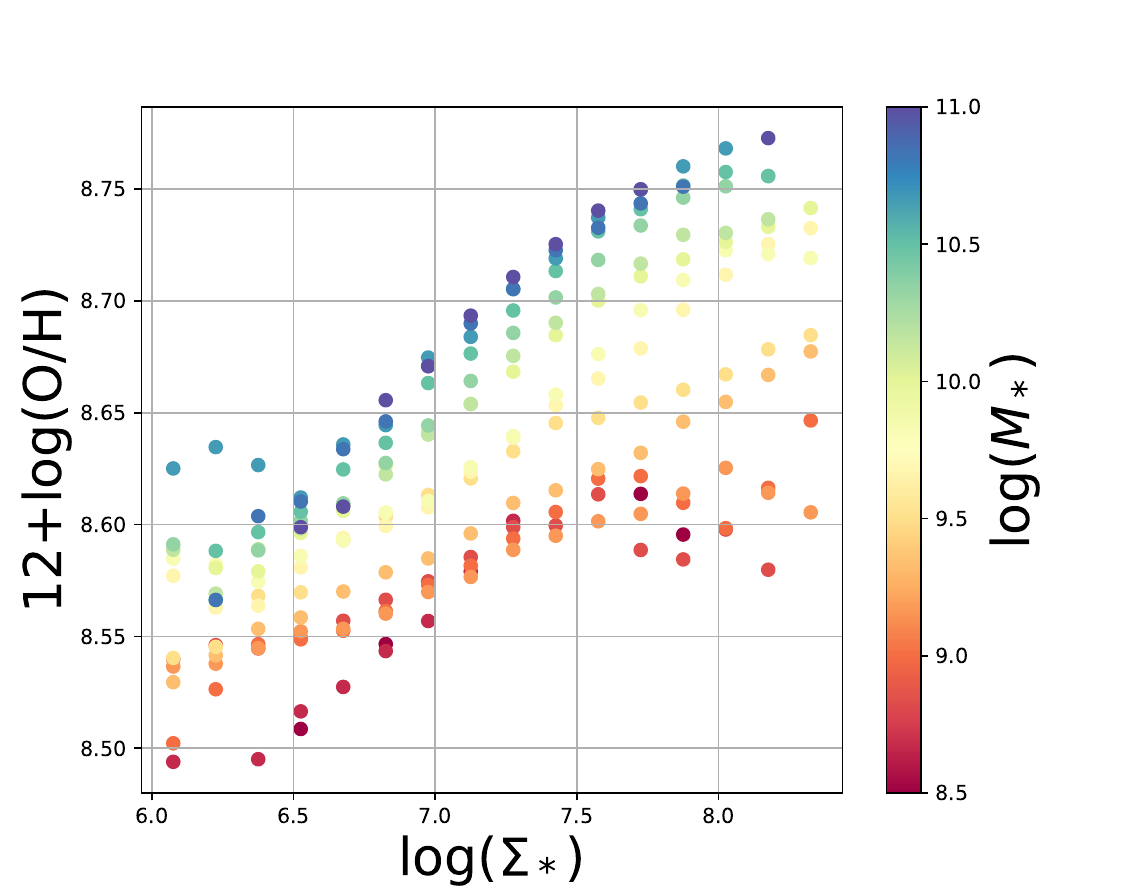}
    \caption{Top: 3D projection of resolved stellar mass, total stellar mass and metallicity. The coloured points show the metallicity in bins of $\Sigma_*$ and $M_*$, colour-coded by $M_*$. The grey points represent the best fit surface to this relation. This shows that there exists a tight relationship between resolved stellar mass, global stellar mass and metallicity. Bottom-left: projection that minimizes the metallicity dispersion. Bottom-right: projection of the surface in the metallicity-$\Sigma_*$ plane, colour-coded by total stellar mass. The 3D surface and colour-coding further illustrate the dependence of the local metalliciy on both local and global stellar mass}
    \label{fig:Mstar_param}
\end{figure*}

\begin{figure}
    \centering
    \includegraphics[width=\columnwidth]{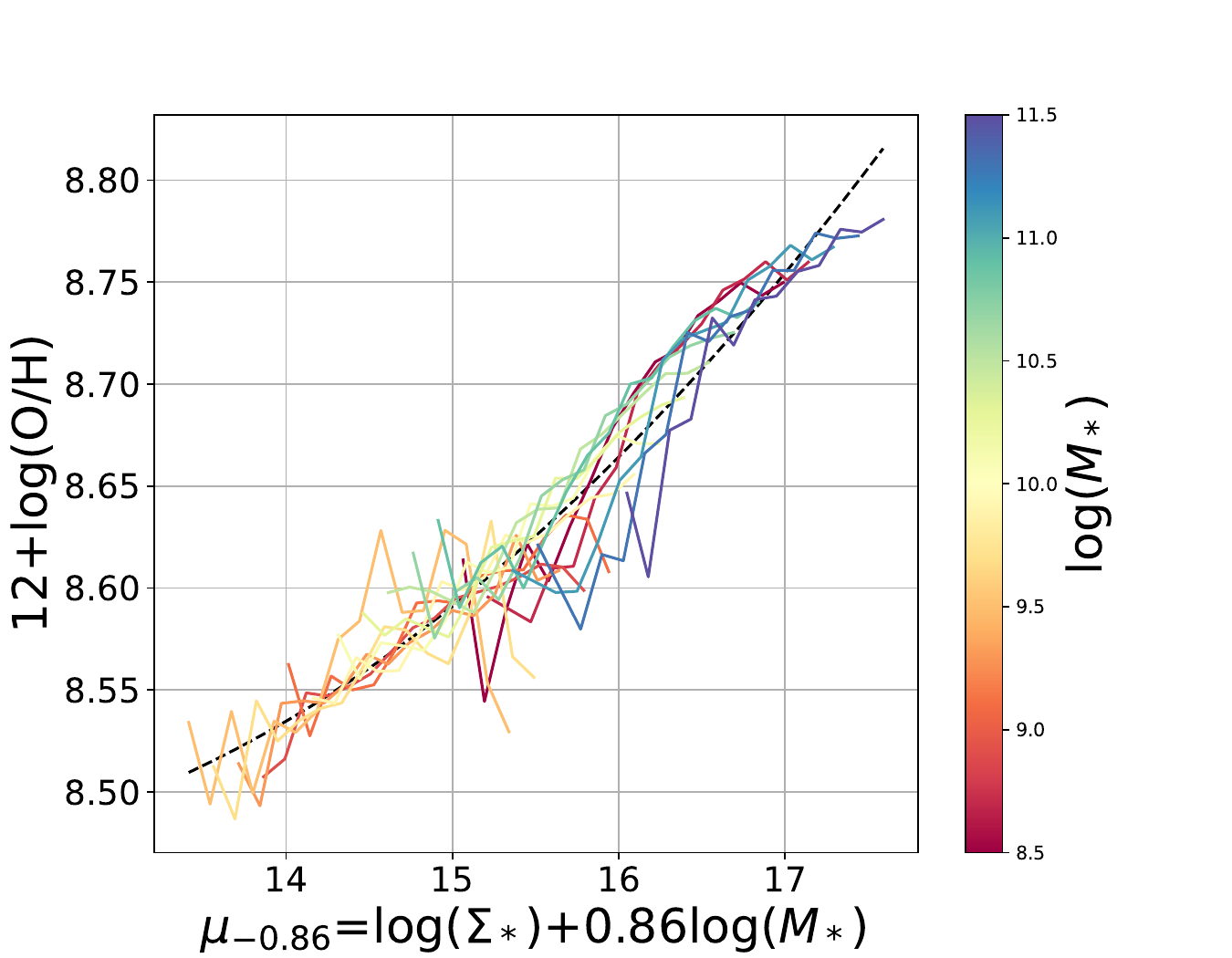}
    \caption{Metallicity against $\mu$=log($\Sigma_*$)-$\beta\,$log($\Sigma_{SFR}$), where $\beta=-0.855$, for tracks of log($M_*$) which minimise the scatter from the best fit relation. The black dashed line is a best fit 2nd order polynomial to the tracks. The tracks generally follow the best fit curve well. Can see how tracks of greater log($M_*$) are offset to higher metallicity showing once again that the resolved metallicity strongly depends on the total stellar mass as well as the resolved stellar mass.}
    \label{fig:mass_alpha_tracks}
\end{figure}

As we have shown that the resolved metallicity depends on both $\Sigma_*$ and $M_*$, we also show this as a 3D surface in Figure \ref{fig:Mstar_param} where the coloured points are bins of $\Sigma_*$-$M_*$, with the mean metallicity per bin, and the grey points are the best-fit surface to the bins. We use the same functional form of surface as in \ref{eq:mirko},
but with $\Sigma _{SFR}$ is replaced by $M_*$:

\begin{equation}
    12 + \rm log(O/H)=Z_0 - \gamma/\phi * \rm log\bigg(1+\bigg(\frac{\Sigma_*}{M_0(M_*)}\bigg)^{-\phi}\bigg)
    \label{eq:mirko_M}
\end{equation}
where $\log{(M_0(M_*))}=m_0+m_1\log{(M_*)}$.
The best fit parameters are  Z$_0$=8.760 ($\pm$ 0.011), $\gamma$=0.092 ($\pm$ 0.002),  $\phi$=3.000 ($\pm$ 1.882), $m_0$= 16.423 ($\pm$ 0.323), $m_1$= -0.814 ($\pm$ 0.029), and $\beta$ -0.855. We also show a projection that minimises the scatter in metallicity (bottom-left) panel, and the simple projection of metallicity-$\Sigma_*$ color-coded by stellar mass.
The 3D plot shows that in general the bins are reasonably well described by this representation, although with difficulties at the lowest total stellar mass end, i.e. log($M_*$/M$_\odot$)<9.5, where the behavior of the bins is no longer well fit by the surface. In light of this we recommend that this parameterisation is only appropriate for galaxies with logarithmic total stellar masses of greater than 9.5.  We can also minimise the scatter in a similar way as for the rFMR surface, only with $\Sigma_{SFR}$ replaced by $M_*$, i.e. we introduce the quantity
\begin{equation}
 \rm   \mu_\beta=log(\Sigma_*) -\, \beta\,log(M_*).
 \label{eq:mu_beta}
\end{equation}

We find that the value of $\beta$ that minimises scatter is $\beta = -0.855$.

Figure \ref{fig:mass_alpha_tracks} shows the tracks of log(M$_*$/M$_\odot$) on the metallicity versus $\mu_{-0.855}$ plot and how their scatter from the best fit curve is minimised in the case of $\beta=-0.855$.
The equation of the second order polynomial best fit is given by
\begin{align}
    \rm 12+log(O/H)=9.522 (\pm 0.326) -0.1890 (\pm 0.042)\mu_{-0.85} \\+0.008 (\pm 0.001)\mu_{-0.85}^2.
\end{align}

\section{The effect of the CO conversion factor}

\begin{figure}
    \centering
    \includegraphics[width=\columnwidth]{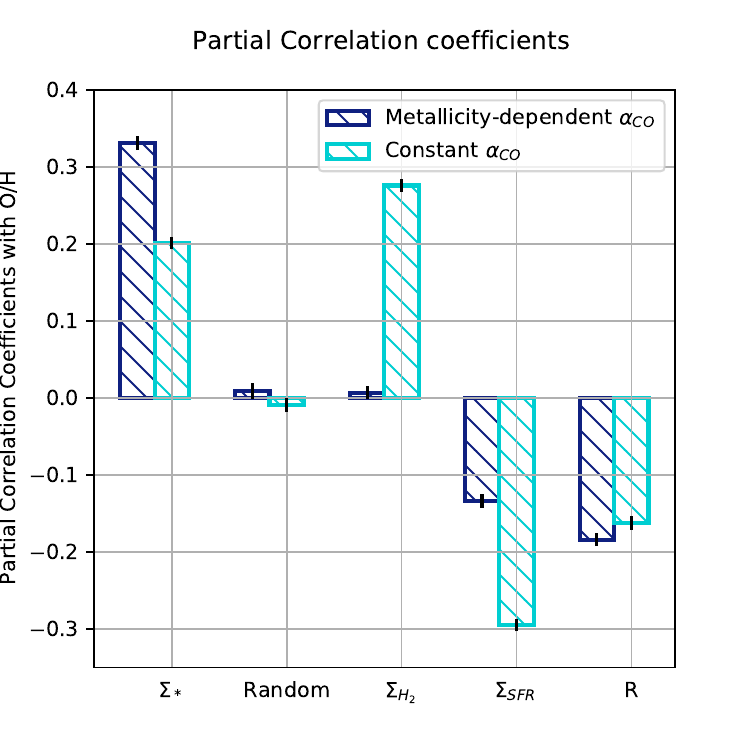}
    \caption{Partial Correlation Coefficients between local metallicity and molecular gas mass surface density ($\Sigma_{H_2}$),  stellar mass surface density ($\Sigma_*$), a uniform random variable (Random),  galactocentric radius (R), and star formation rate surface density ($\Sigma_{SFR}$). Dark blue bars are for a metallicity dependent CO-to-H2 conversion factor, which we used as our baseline, and for which there is general consensus, while the light blue bars are a test for the extreme (unrealistic) case of constant CO-to-H2 conversion factor.  We see that there are major differences in the PCC results when the unrealistic conversion factor is applied.}
    \label{fig:pcc_app}
\end{figure}

\begin{figure}
    \centering
    \includegraphics[width=\columnwidth]{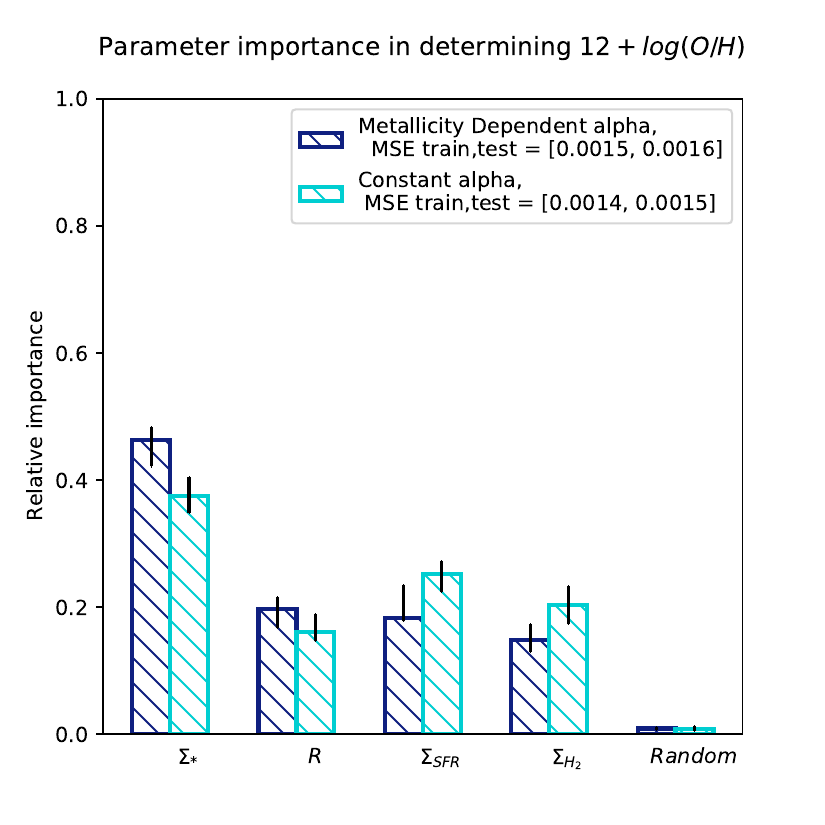}
    \caption{Random Forest regression parameter importance for determining the resolved metallicity in the ALMaQUEST sample. Parameters included are:  stellar mass surface density ($\Sigma_*$),  star formation rate surface density ($\Sigma_{SFR}$),  molecular gas mass surface density ($\Sigma_{H_2}$), galactocentric radius (R), and a random uniform variable (Random). Dark blue bars are for a metallicity dependent CO-to-H2 conversion factor, for which there is general consensus, while the light blue bars are a test for an extreme (unrealistic) case of constant CO-to-H2 conversion factor. Interestingly, in comparison to the PCC, the RF does not appear to be as affected by the unrealistic conversion factor. }
    \label{fig:RF_alma_app}
\end{figure}

In section \ref{sec:alma_rf} we explored the resolved metallicity's dependence on molecular gas mass surface density using partial correlation coefficients and random forest regression. To do this we assumed a metallicity-dependent conversion factor \citep[as given in][]{Accurso2017MNRAS.470.4750A, Sun2020ApJ...892..148S}. Here,
 as a sanity check, we explore the results for an (implausible) constant conversion factor. 
 We know that this assumption is incorrect, as shown by multiple studies \citep[][]{Bolatto2013,Accurso2017MNRAS.470.4750A}, but it is anyway useful to explore its effect.
The light blue bars in Figures \ref{fig:pcc_app} and \ref{fig:RF_alma_app} show the case of adopting a constant (metallicity independent) CO-to-H2 conversion factor.  Naively, one would expect that removing the dependence of $\alpha _{CO}$ on the metallicity would further decrease the correlation between $\Sigma _{H_2}$ and metallicity, instead we observe the opposite effect: in the RF analysis $\Sigma _{H_2}$ becomes the third most important parameter and in the PCC analysis becomes the most important parameter, which also makes $\Sigma _*$ the least important parameter (in contrast with all other results, including those of the full MaNGA sample). Moreover, the sign of the strong correlation (strongly positive) is {\it opposite} of what is expected from the FMR. This result can be understood in terms of $\alpha _{CO}$ having in reality a steep dependence on the metallicity \citep{Sun2020ApJ...892..148S,Bolatto2013} and, therefore, when this dependence is artificially removed by assuming a constant $\alpha _{CO}$, then $\Sigma _{H_2}$ tries to compensate by artificially becoming strongly dependent on the metallicity.

\section{Effect of measurement uncertainty}

\label{sec:measurementuncertainty}

\begin{figure}
    \centering
    \includegraphics[width=\columnwidth]{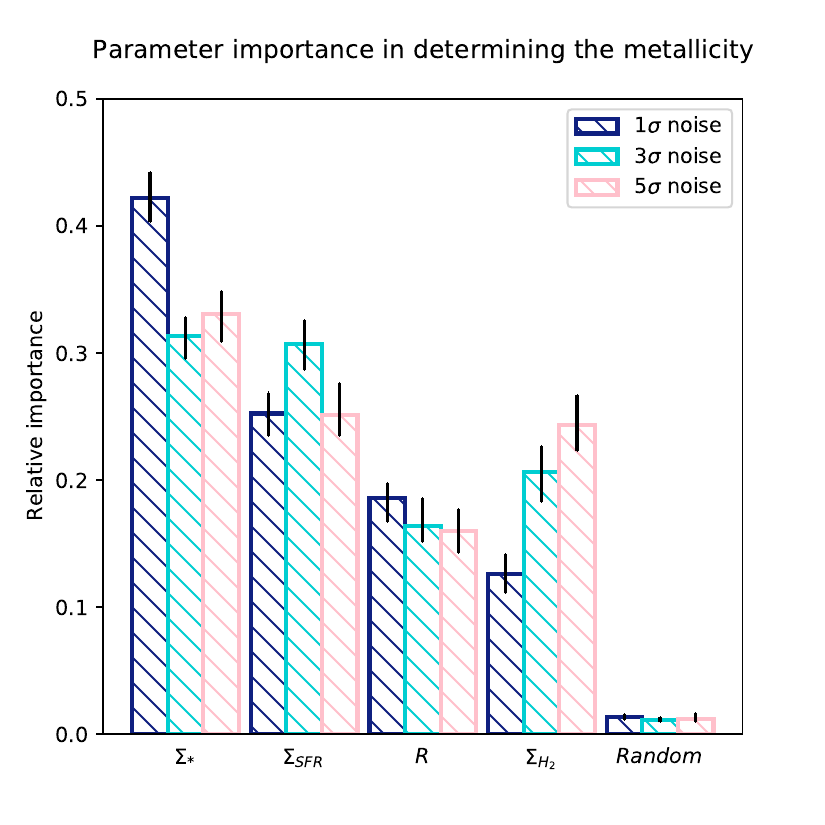}
    \caption{A bar-chart showing the random forest regression parameter importances in determining the metallicity. The parameters evaluated are the same as those in Figure \ref{fig:RF_alma}. Each coloured bar corresponds to an increasing amount of noise added to $\Sigma_*$, $\Sigma_{SFR}$ and R. The figure shows that even after 5$\sigma$ of noise is added to $\Sigma_{SFR}$, $\Sigma_{H_2}$ is still not responsible for driving the metallicity. This means that the random forest is not being biased by varying measurement uncertainties.  }
    \label{diff_measurement_uncert_test}
\end{figure}

In section \ref{sec:alma_rf} we found from the random forest regression that $\Sigma_{SFR}$ was more important than $\Sigma_{H_2}$ in determining the metallicity. However, it is important to test that this result is not a product of greater measurement uncertainty on $\Sigma_{H_2}$. To do this we perform a differential measurement uncertainty test following the method in \cite{Baker2022MNRAS.510.3622B}. We add increasing multiples of Gaussian random noise to $\Sigma_*$, $\Sigma_{SFR}$ and R (the galactocentric radius). We select the uncertainty on $\Sigma_{H_2}$ to be $\sim$0.22 dex which is the scatter of the resolved Schmidt-Kennicutt relation in \cite{Baker2022MNRAS.510.3622B}. 

Figure \ref{diff_measurement_uncert_test} shows the results of this test. The figure uses exactly the same method as Figure \ref{fig:RF_alma}, but with increasing amounts of noise added to $\Sigma_*$, $\Sigma_{SFR}$ and R. This figure shows that even after $5\sigma$ of Gaussian random noise is added, $\Sigma_{H_2}$ is still no more important than $\Sigma_{SFR}$ in determining the metallicity. Hence, the previous result is robust to the varying levels of measurement uncertainty in the parameters evaluated.


\bsp	
\label{lastpage}
\end{document}